\documentclass[apjl,twocolumn]{emulateapj}

\usepackage[backref,breaklinks,colorlinks,citecolor=blue]{hyperref}        
\usepackage[all]{hypcap}

\usepackage{amsmath}
\usepackage{amssymb}
\usepackage{graphicx}
\usepackage{color}
\usepackage{natbib}
\usepackage{xspace}
\usepackage{nicefrac}
\usepackage[dvipsnames]{xcolor}
\newcommand\myshade{85}
\usepackage[T1]{fontenc}
\usepackage{lmodern}
 	
\colorlet{mycitecolor}{Turquoise}
\colorlet{mylinkcolor}{Turquoise}

\hypersetup{
  citecolor  = mycitecolor!\myshade!black,
  linkcolor  = mylinkcolor!\myshade!black,
  colorlinks = true,
}

\newcommand{\Gyr}{\ensuremath{\,\mathrm{Gyr}}\xspace}

\def\Msun{{\rm M}_{\odot}} 

\newcommand{\kms}{\ensuremath{\,\rm{km}\,\rm{s}^{-1}}}

\def\apjl{Astrophysical Journal}             

\def\apjs{ApJS}              
\def\aap{A\&A}                

\def\be{\begin{equation}}
\def\ee{\end{equation}}
\def\ba{\begin{eqnarray}}
\def\ea{\end{eqnarray}}

\submitted{Submitted in original form: Jan 1, 2016; Accepted for publication in MNRAS: Feb. 12, 2016. }

\begin{document}

%

\title{Merging binary black holes formed through \\ chemically homogeneous evolution in short-period stellar binaries}
\shorttitle{Case M binary black holes}

\author{ I. Mandel\altaffilmark{1,*} \& S. E. de Mink\altaffilmark{2,*}
} 

 \affil{
     $^{1}$School of Physics and Astronomy, University of Birmingham, Birmingham B15 2TT, UK
               (IMandel@star.sr.bham.ac.uk)\\
               $^{2}$Anton Pannenkoek Institute for Astronomy, University of
Amsterdam, 1090 GE Amsterdam, The Netherlands (S.E.deMink@uva.nl)\\
$^{*}$Both authors contributed equally to this work. 
         }




\begin{abstract}
We explore a newly proposed channel to create binary black holes of stellar origin. This scenario applies to massive, tight binaries where mixing induced by rotation and tides transports the products of hydrogen burning throughout the stellar envelopes. This slowly enriches the entire star with helium, preventing the build-up of an internal chemical gradient. The stars remain compact as they evolve nearly chemically homogeneously, eventually forming two black holes, which, we estimate, typically merge 4--11 Gyr after formation.  Like other proposed channels, this evolutionary pathway suffers from significant theoretical uncertainties, but could be constrained in the near future by data from advanced ground-based gravitational-wave detectors.
We perform Monte Carlo simulations of the expected merger rate over cosmic time to explore the implications and uncertainties. Our default model for this channel yields a local binary black hole merger rate of about $10$ Gpc$^{-3}$ yr$^{-1}$ at redshift $z=0$, peaking at twice this rate at $z=0.5$.  This means that this channel is competitive, in terms of expected rates, with the conventional formation scenarios that involve a common-envelope phase during isolated binary evolution or dynamical interaction in a dense cluster.  The events from this channel may be distinguished by the preference for nearly equal-mass components and high masses, with typical total masses between 50 and 110\,$\Msun$. Unlike the conventional isolated binary evolution scenario that involves shrinkage of the orbit during a common-envelope phase, short time delays are unlikely for this channel, implying that we do not expect mergers at high redshift.  
\end{abstract}


\section{Introduction}
Ground-based gravitational-wave detectors \citep{AdvLIGO, AdvVirgo} are capable of observing gravitational waves emitted during mergers of compact-object binaries composed of neutron stars and black holes of stellar origin \citep{ratesdoc}.  The anticipated increasing sensitivity of these detectors during the rest of the decade \citep{scenarios} motivates a careful reconsideration of the physical processes that affect the evolution of massive stars in binary systems as progenitors of double compact mergers \citep[e.g., ][]{Belczynski:2015}. One of these processes is stellar rotation.  Rotation can trigger mixing processes in layers of stars that would otherwise be stable \citep[e.g.,][]{Endal+1976}, affects mass loss by stellar winds \citep{Georgy+2011}, and thus influences the evolution of the progenitors of neutron stars and stellar-mass black holes.  

The inclusion of the effects of stellar rotation is one of the major recent developments in theoretical modeling of the evolution of massive stars. Rotation has become a standard ingredient in various evolutionary codes \citep[e.g.,][]{Maeder+2000a,  Heger+2000, Potter:2012, Paxton+2013} and extensive model grids of single stars are now becoming available \citep[e.g.,][]{Brott+2011,Ekstrom+2012, Kohler+2015}.  Several studies of the effects of rotation on massive stars in binary systems have been undertaken \citep[e.g.,][]{Cantiello+2007, de-Mink+2009a, de-Mink+2013, Song+2013,Song+2015}, but the implications for the formation of double compact objects have not yet been fully considered.

One of the most intriguing predictions of the rotating models is the possibility of the so-called \emph{chemically homogeneous evolution}, first described by \citet{Maeder1987}.  A very rapidly rotating star may experience mixing, which allows the star to transport material from the hydrogen-rich envelope into the central burning regions and vice versa.  If these processes are efficient, the build-up of internal chemical gradients is prevented and the star evolves (quasi) chemically homogeneously.  Such stars slowly contract as their envelopes become more and more helium rich, evolving to become hotter, more luminous and more compact.  Theoretical models indicate that this evolutionary pathway is favored at low metallicity \citep{Yoon+2005, Yoon+2006}. 

\citet{de-Mink+2009a} considered the possibility and implications of chemically homogeneous evolution in near contact binary systems. In such systems tides force the stars to spin rapidly, synchronized with the orbital revolution.   Their binary models \citep[and earlier models in][]{de-Mink+2008c} show that the stellar spins achieved in very tight binary systems are sufficient to reach the conditions of chemically homogeneous evolution.  They argue that chemically homogeneous evolution may be further favored in such systems, because of additional mixing processes that are expected but not yet accounted for in the models, for example those arising from tidal deformation. 

Binaries composed of two chemically homogeneously evolving stars, which shrink inside their Roche lobes as they gradually convert nearly all their hydrogen into helium, therefore proceed on a relatively simple evolutionary pathway, avoiding the complexities of mass transfer, including common-envelope phases. This evolutionary scenario predicts the formation of two massive helium stars that may eventually collapse to form two stellar-mass black holes. 

There are significant uncertainties associated with the physical processes that determine the formation of double compact objects through this channel, which is also true for all other proposed formation channels. The classical isolated binary formation scenario involves one or more phases of common-envelope evolution. The key uncertainties lie in the treatment of Roche lobe overflow and the ejection of the common envelope.  For this new scenario the main uncertainties arise from the mixing processes, which at present cannot be treated self consistently in the 1D evolutionary models, and the effects of stellar wind-driven mass loss on the binary orbit.  So far, the massive overcontact binary VFTS~352 \citep{Almeida+2015} appears to be the most promising example of a system experiencing enhanced mixing, but solid observational evidence for this new scenario is missing. Constraints from electromagnetic observations, i.e., spectroscopic and photometric campaigns, are challenging due to the rarity of this channel and its preference for low metallicity. Gravitational-wave observations can help to constrain the physics of massive binary evolution, including through the chemically homogeneous formation channel, by probing the mergers of evolutionary end products or, in the case of no detections, by providing upper limits on the merger rates \citep[e.g.,][]{MandelOShaughnessy:2010,Belczynski:2015,Stevenson:2015}.

In this paper, we explore the implications of the chemically homogeneous evolution channel for the formation of binary black holes and their merger rates.  We estimate the cosmic and local merger rates and the typical properties of the merging binaries with a Monte Carlo simulation. We show that the binary black holes formed through this channel typically merge in 4--11 Gyr in our default simulations.  The expected merger rates are competitive with other proposed pathways for the formation of binary black holes, with default-model merger rates of $\sim 10$ Gpc$^{-3}$ yr$^{-1}$ at $z=0$, peaking at $\sim 20$ Gpc$^{-3}$ yr$^{-1}$ at $z\sim 0.5$.  We further discuss the testable features of this channel, including high binary masses (total masses of $\sim 50$ -- $100\, \Msun$ for merging binary black holes), a preference for equal masses (component masses differ by no more than a factor of two at merger), the likelihood of aligned spins and the lack of very short delay times.  

The paper is organized as follows. We discuss rapid rotation in binary systems and the ensuing chemically homogeneous evolution in \autoref{sec:homog}.   We provide a back-of-the-envelope estimate for the merger rate of binaries produced through this channel in \autoref{sec:BOTE}.  We list the input assumptions in \autoref{sec:inputs} and describe the setup of the Monte Carlo simulation in \autoref{sec:MC}.  The results for our default simulation are given in \autoref{sec:results}, while alternative models which delineate the theoretical uncertainties are analyzed in \autoref{sec:uncertain}.  We conclude with a summary in \autoref{sec:summary}.

\section{Chemically homogeneous evolution in binary systems}\label{sec:homog}

\subsection {Mixing processes in non-rotating stars}

During their first phase of evolution massive stars fuse hydrogen into helium in the center through the CNO-cycle. Their central regions are unstable against convection. Convective mixing efficiently supplies fresh hydrogen to the very center where the temperatures are high enough for nuclear burning.  The envelope is stable against convection, apart from very small regions near the surface \citep{Maeder1980, Cantiello+2009}. The elements produced in the center cannot reach the stellar surface, unless the envelope is removed, for example by stellar wind-driven mass loss.   

Mixing beyond the boundary of the convective core is possible, for example when convective shells penetrate into the  radiative layer above due to their inertia.  This process, generally referred to as overshooting, will mix a limited region above the convective core. The extent of this region is typically parametrized in units of the local pressure scale height. Calibrations against observations point to values between 0.1 and 0.6 \citep{Pols+1997,Schroder+1997, Ribas+2000, Claret2007a, Brott+2011a, Stancliffe+2015}.  

Overshooting can increase the mass of the stellar core; however, the core-envelope structure, i.e., the steep gradient in composition and density, remains intact.  The existence of this transition plays a key role in the evolution of the stellar structure. The core contracts and becomes denser as hydrogen is converted into helium. To remain in hydrostatic equilibrium the envelope responds by expanding, which is often referred to as the mirror effect.  By how much the star expands depends on details in the chemical profile near the core, the metallicity and mass loss, but the general trend of stellar envelopes to expand as the stellar core contracts is observed at all evolutionary phases.  However, this behavior is absent in very well mixed stars that lack a core-envelope structure, as we will discuss below.  

\subsection {Stellar rotation and mixing processes in rotating stars}

Young massive stars are observed to rotate with a wide range of (projected) rotational velocities \citep[e.g.,][]{Penny+2009, Dufton+2013, Ramirez-Agudelo+2015}. The majority spins at moderate rates, corresponding to about 10-20\% of the Keplerian rate. However, the distribution shows a large spread including systems that rotate close to the Keplerian (``break up'') angular frequency \citep[e.g.,][]{Dufton+2011, Ramirez-Agudelo+2013}. 

As first shown already by \citet{von-Zeipel1924,von-Zeipel1924a} a rotating star cannot be in hydrostatic and radiative thermal equilibrium at the same time because surfaces of constant temperature and constant pressure do not coincide. As a result of this large-scale meridional circulations develop \citep{Eddington1925, Sweet1950}. These circulations can cause mixing in the radiative envelopes of massive stars that would otherwise be stable against mixing.   

In addition, as rotating stars evolve their interior layers contract and tend to spin up while the outer layers normally expand.  This naturally leads to internal shear, which can also lead to mixing of layers that are otherwise stable.   Pioneering work comes from \citet{Endal+1978}, who provided order-of-magnitude estimates for the efficiencies of various instabilities, and performed time-dependent calculations of the evolution of rotating massive stars.   The dynamical shear instability occurs when the energy that can be gained from the shear flow becomes comparable to the work that has to be done against the gravitational potential for the adiabatic turnover of a mass element.  This criterion can be relaxed by allowing for thermal adjustments. In this case the process operates on a thermal timescale, and is referred to as the secular shear instability \citep{Endal+1978,Heger+2000}.

Rotational mixing was originally invoked to explain the surface enrichment of some massive main-sequence stars with the products from hydrogen burning, in particular nitrogen \citep[e.g.,][]{Maeder+2000a,Maeder2000}. The inclusion of the effects of rotation has become standard in the state-of-the art detailed evolutionary calculations. However, quantitative predictions differ, depending on how these effects are accounted for \citep[e.g.,][]{Brott+2011,Ekstrom+2012, Potter:2012, Paxton+2013} .

\subsection {Chemically homogeneous evolution in single stars}\label{CHEsingle}

One of the most intriguing predictions of the rotating models is the possibility of the so-called chemically homogeneous evolution.  As first shown by \citet{Maeder1987}, the internal mixing processes induced by rotation may lead to a bifurcation in the evolutionary paths of massive stars.  

Slowly rotating stars build a strong internal composition gradient between their increasingly helium-rich convective core and their hydrogen-rich envelope. As the core contracts the star adapts itself to the composition changes by expanding the envelope in order to maintain hydrostatic and thermal equilibrium. This leads to the typical red-ward evolutionary expansion that characterizes the main evolutionary stages of all stars.  

In contrast, a very rapidly rotating star may experience mixing which allows the star to transport material from the hydrogen-rich envelope into the central burning regions and vice versa.  If rotationally-induced instabilities become so efficient that they prevent the build-up of a chemical gradient which separates the core from the envelope, the star will evolve quasi chemically homogeneously.  Such stars become brighter and bluer as they evolve, with their radii staying nearly constant during the main sequence, very close to fully homogeneous evolutionary tracks. When hydrogen is exhausted in the center (and throughout most of the envelope) the star contracts towards the helium main sequence. This evolutionary path leads to the production of very massive helium stars \citep{Yoon+2006}, which convert a larger fraction of their initial mass into helium than non-rotating stars with the same initial mass and metallicity.  

This peculiar evolutionary path gained renewed attention when \citet{Yoon+2005} and \citet{Woosley+2006} proposed it as a way to produce the rapidly spinning massive helium progenitors of long gamma-ray bursts in the collapsar scenario \citep{Woosley1993}. \citet{Yoon+2006} further studied the parameter space by presenting models for different masses $10-60\Msun$,  a range of rotation rates and metallicities of $Z=0.004, 0.002, 0.001$ and $10^{-5}$, where $Z$ is the combined mass fraction of all elements heavier than helium.  Chemically homogeneous evolution was found at all metallicities considered, see Fig.~3 in \citet{Yoon+2006}.  They find that more massive stars are more prone to evolve chemically homogeneously. They attribute this to the fact that the entropy barrier becomes weakened in more massive stars due to the increased role of radiation pressure. In addition they note that the ratio of the thermal to the nuclear timescale decreases with mass.  As a result they find that the critical minimum rotation rate for chemically homogeneous evolution is 20--30\% of the Keplerian rotation rate for stars around $60\Msun$.  It should be noted that these values are model dependent.

\citet{Brott+2011} and \citet{Kohler+2015}  further explored a larger range of masses and higher metallicities using models where the efficiency of rotational mixing is calibrated against the observed nitrogen abundances of early B type stars in the Large Magellanic Cloud. Their grid covers metallicities of  $Z=0.0088, 0.0047$ and $0.0021$.   They find chemically homogeneous evolution in their two lowest metallicity models, for stars more massive than $20\Msun$, where the initial rotational velocity at which a star evolves homogeneously decreases with increasing initial mass.  The grid computed by \citet{Szecsi+2015} explores $Z = 0.0002$ and finds chemically homogeneous evolution down to $9\Msun$, the lowest mass explored in their grid.  At higher metallicity, wind-driven mass loss becomes more important; the associated angular momentum loss spins down the stars, shutting off rotational mixing. 

We pause here to emphasize that the model predictions should be interpreted with ample caution. It is clear that massive stars rotate and the models accounting for rotation have been successful in explaining various observations \citep[][and below]{Maeder+2000a}. However, it remains unclear if the approximate treatment of rotational mixing in the 1D stellar evolutionary models provides an adequate prescription of the complex processes. It should also be noted that the direct comparison of trends between the measured projected rotation rates and surface nitrogen abundances has raised questions concerning rotational mixing \citep{Hunter+2008a,Maeder+2009, Brott+2011a}.

\subsection {Observational evidence for the existence of stars evolving chemically homogeneously\label{observations}}

Several studies presented observational clues that could possibly be interpreted as evidence in favor of chemically homogeneous evolution, although no hard evidence exists at present.  Chemically homogeneous evolution is challenging to investigate observationally because of the rarity of rapidly rotating massive stars and the difficulty of observing very metal-poor environments.  

\citet{Martins+2013} perform a spectroscopic analysis of several very hot hydrogen-rich Wolf-Rayet stars and conclude these objects are consistent with the predictions of chemically homogeneous evolution.  More recently, \citet{Almeida+2015} presented an analysis of the very massive over-contact binary system VFTS 352, part of the VLT-FLAMES Tarantula Survey of Massive Stars \citep{Evans+2011}. The system consists of stars with mass $28.63\pm0.30\Msun$ and  $28.85\pm0.30\Msun$ in a 1.12 day orbit.  It is the hottest massive stellar binary system known to date. The stars in this system are found to be too hot and compact for their dynamically inferred masses.  If these parameters are derived correctly, they defy the predictions of non-rotating models. Enhanced mixing processes provide a natural explanation for this system.  The stars in this system are two of the best candidates identified so far for undergoing chemically homogeneous evolution.

Further hints come from the integrated light  of Lyman-break galaxies.  \citet{Eldridge+2012} compare their population synthesis models with the observed strengths of He II{$\lambda$}1640 {\AA} and C IV{$\lambda$}1548, 1551 {\AA} spectral lines of Lyman-break galaxy spectra at $z {\sim} 2-3$.  They can only fit the spectra of the lowest-metallicity sources by boosting the He II emission line by including chemically homogeneously evolving stars produced in binary systems.  \citet{Stanway+2014} draw the same conclusion based on the  [OIII]/H emission line ratios in low-mass star-forming galaxies at subsolar metallicity. Also, \citet{Szecsi+2015} argue that chemically-homogeneously evolving stars may provide an explanation for the He II ionizing photon flux in I Zw 18 and other low-metallicity He II galaxies.

These studies cannot be considered as satisfactory proof of the chemically homogeneous evolution scenario, but justify speculating about its possible implications with the aim of identifying further opportunities to test this scenario.  

\subsection {Chemically homogeneous evolution in binary systems}

\begin{figure}\center
  \includegraphics[width=\columnwidth]{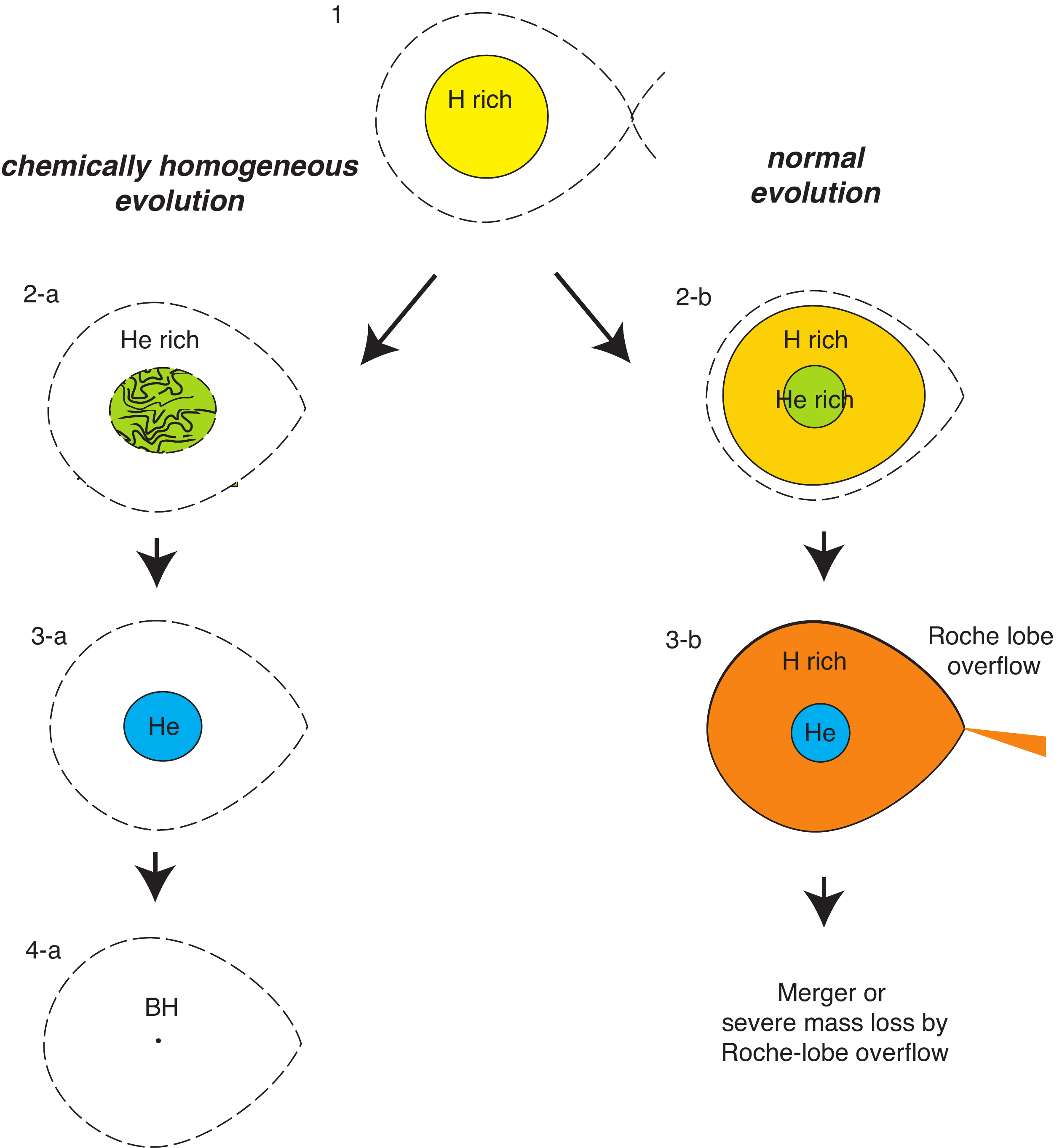}
  \caption{A schematic representation of the implications of  ``normal'' versus chemically homogeneous evolution in a close binary system.  The effects of enhanced mixing cause the star to shrink inside its Roche lobe instead of expanding, and avoid a large amount of mass loss.  This evolutionary path can, in principle, lead to the formation of massive stellar black holes in a close binary system.    Figure adapted from \citet{de-Mink+2008c}.    \label{cartoon}}
\end{figure}

High stellar spins can be achieved in binary systems as a result of spin-up by mass accretion \citep{Packet1981, Cantiello+2007, de-Mink+2013} or tidal spin-up in very close binary systems \citep{Zahn1989, Izzard+2004b, Detmers+2008, de-Mink+2009a}.  In the latter systems, when tides synchronize the stellar rotation rates with the orbital revolution, the conditions for chemically homogeneous evolution can be reached.   \citet{de-Mink+2008c, de-Mink+2009a} demonstrated this possibility with binary evolutionary calculations adopting the same assumptions as \citet{Yoon+2006} and \citet{Brott+2011}, respectively, for the rotationally induced mixing processes.  

This can lead to surprising effects.  The classical models predict that the two stars in very close binaries come in contact soon after the onset of Roche-lobe overflow and are expected to merge.  The possibility of chemically homogeneous evolution changes this classic picture, leading to a type of evolution referred to as Case M by \citet{de-Mink+2009a} to emphasize the role of mixing; this is illustrated in \autoref{cartoon}. The two stars slowly shrink inside their Roche lobe as they become more and more helium rich. Over the course of the main sequence they stay within but close to their Roche lobes.  When no more hydrogen is left in the center, the stars fully contract to form a massive double helium star binary, without ever overfilling their Roche lobes and initiating mass transfer, preventing both severe mass loss and possible merger. 

The rotational rates required for chemically homogeneous evolution, 20-30\% of the Keplerian velocity (see \autoref{CHEsingle}), can be achieved in very close tidally locked binary systems.  In a tidally locked binary system, where the nearly equal-mass stars are close to filling their Roche lobe, synchronized spins correspond to about a third of the Keplerian rotational velocity.  This means that there should be a small parameter-space window for chemically homogeneous evolution in tidally locked binary systems. This assumes that the mixing processes in tidally locked binaries are at least as efficient as they are in single stars.  Detailed simulations of such systems were presented initially by \citet{de-Mink+2009a} and later by \citet{Song+2013,Song+2015, Marchant+2016}.  

\begin{figure}\center
  \includegraphics[width=\columnwidth]{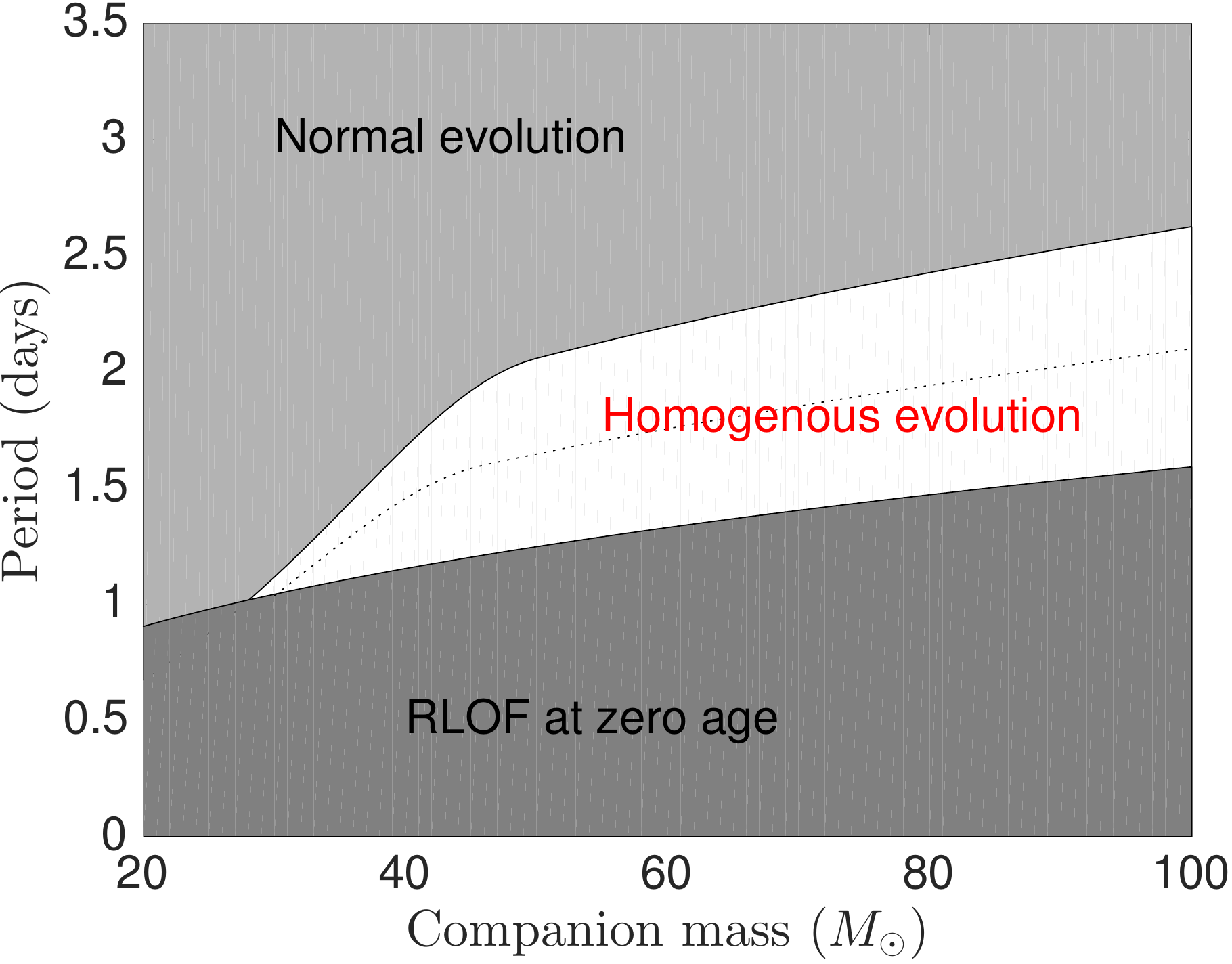}
  \caption{The window for chemically homogeneous evolution in close tidally locked equal-mass binary systems, where the stellar spin period equals the orbital period. The lower part of the diagram is excluded due to the finite size of the stars at zero age. In the upper part of the diagram the stellar components do not rotate rapidly enough to evolve chemically homogeneously during the central hydrogen burning phase according to the models of \citet{Yoon+2006}.  Stars in the intermediate window, with masses $\gtrsim 40\Msun$ and periods of $\sim 1.5$ -- $2.5$ days, may undergo this type of evolution and avoid Roche-lobe overflow entirely.   Only binaries below the dotted line satisfy the more stringent threshold on chemically homogeneous evolution introduced in \autoref{altCHE}.
\label{parspace}}
\end{figure}

In  Fig.~\ref{parspace} we visualize the parameter space in a diagram similar to the one first presented in \citet{de-Mink+2008c}.  The short-period systems would already overflow their Roche lobes at zero age, and are therefore excluded.  For wide-period systems, tidal synchronization results in spin periods that are too low for chemically homogeneous evolution, and we expect the stars to evolve normally.  We find a small window for stars with masses $\gtrsim 40\Msun$ and orbital periods between $\sim 1.5$ and $\sim 2.5$ days, which permits chemically homogeneous evolution. 

\subsection {The role of stellar winds: mass loss and angular momentum loss}\label{winds}

At the metallicities of interest for this channel, $Z \lesssim 0.004$, the radiatively driven winds are strongly reduced as predicted by \citet{Vink+2000,Vink+2001} and empirically verified by \citet{Mokiem+2007a}.  However, given the brightness and high temperatures that the homogeneous stars reach, stellar wind mass loss cannot be neglected. 

Stellar wind mass loss and its associated uncertainties will affect our results in several ways. Most importantly wind mass loss reduces the masses and affects the final orbit. Both will in turn affect the predicted time needed for the final merger as well as the final masses of the compact objects. If the orbit widens too much, the synchronized rotation rate may become too low for chemically homogeneous evolution.

The models by \citet{Yoon+2006} overestimated the amount of mass loss. The authors assumed that self-enrichment of the surface with the star's own burning products efficiently boosts the wind.  However, the main driver for the wind is iron, which the star cannot produce during its evolution \citep{Vink+2005, Maeder+2012}.

How mass loss affects the orbit depends on whether the stellar wind will have time to interact with the system.  In the extreme limit for very fast isotropic mass loss, the wind has no time to interact with the system, and will simply take away the specific orbital angular momentum of the mass losing star. This mode of mass loss is referred to as the Jeans mode; the change $da$ in the binary's semimajor axis $a$ following a small change in the mass of the binary $dm \ll m_{1,2}$ where $m_{1,2}$ are the component masses is given by
\begin{equation}\label{dadm} 
da = a \frac{|dm|}{m_1+m_2}\ .
\end{equation}

In reality, however,  we are not fully in this idealized regime.  The merger rate is dominated by massive systems which have typical orbital velocities of components relative to each other of around $v_{\rm orb} \sim 800$\kms (corresponding to a system with $60 M_\odot$ components in a 2-day orbit) and higher.  This is comparable to the expected wind velocities.  Terminal wind speeds for typical Galactic Wolf-Rayet stars are  $v_{\infty} \sim$ 800--1200 \kms\ \citep{Vink+2005}.  At lower metallicity, the wind speeds are expected to be reduced, but the effect is fairly weak, $v_{\infty} \propto Z^{0.13}$  \citep[][and J.~S.~Vink private communication]{Leitherer+1992}, implying a reduction in wind speeds by 10-20\% for $Z\gtrsim 0.2 Z_\odot$.   Furthermore, our binary systems are so tight that the companion resides inside the wind acceleration zone and the winds will not yet have reached their terminal speeds.   This means that we are in the regime where simulations by  \citet{Brookshaw+1993} indicate that the widening is significantly less than expected in the Jeans mode of mass loss, and wind interactions may even harden the binary.

In our standard simulation we assume mass loss through non-interacting winds ($v_{\rm wind} \gg v_{\rm orb}$). We consider the effect of slower winds when estimating the model uncertainties.

\subsection{Context of binary population synthesis models}

The predictions for the merger rates of double neutron stars (NS) can be derived semi-empirically using the observed binary pulsars in our Galaxy \citep[e.g.,][]{Phinney1991,Narayan:1991,Kalogera+2004}.  For double compact object binaries involving a black hole of stellar origin, we must fully rely on model predictions, since we lack direct observational evidence of such systems \citep[see][for a review]{MandelOShaughnessy:2010}. We do, however, have observations of BH systems with a stellar companion that provide indirect constraints for binary black hole \citep[e.g.,][]{Bulik+2011} and black hole -- neutron star systems \citep[e.g.,][]{CygnusX3:2012,Grudzinska+2015}. 

The majority of predictions rely on Monte Carlo codes which use approximate but rapid recipes to simulate the evolution of the progenitor systems.  Such approximations allow these ``population synthesis'' codes to explore the wide parameters space that is inherent to the evolution of binary systems.  The price paid for the required computational efficiency is a set of simplifications; for example, stellar structure is not generally evolved directly, but relies on a set of models such as those provided by 
\citet{Hurley+2000,Hurley+2002}, which in turn rely on one original grid of non-rotating stellar models \citep{Pols+1998}.  Groups that have estimated compact binary merger rates in the past include \citet{Lipunov+1997,Bethe+1998,Bloom+1999,De-Donder+2004,Grishchuk+2001,
Nelemans:2003,Voss+2003,Pfahl+2005,Dewi:2006,OShaughnessy:2008,Mennekens+2014,Dominik+2015,de-Mink+2015} \citep[there is also an independent dynamical formation channel in dense stellar environments which we do not discuss here; see][for references]{ratesdoc,Rodriguez:2015}.

These simplified rapid codes \citep[see][for a review]{PostnovYungelson:2014} may still be appropriate for evolving the majority of massive binaries.  However, they do not account for rare minority channels, such as chemically homogeneous evolution, that result from physical processes not accounted for in the original physical models used as input into population synthesis recipes.  Double compact mergers may primarily be the result of such rare minority channels.

\section {A back-of-the-envelope estimate}\label{sec:BOTE}

In the following sections, we describe a Monte Carlo simulation to estimate the rate of mergers and properties of binary black holes evolving through the chemically homogeneous evolution channel.  Here, we carry out a very crude back-of-the-envelope estimate.  This estimate can be viewed as an order-of-magnitude sanity check on the results of the Monte Carlo simulations described in the following sections.

One possible way to proceed with the estimate is to consider a Drake-like equation.  The rate of local mergers per unit volume per unit time is given by:
\be \label{drake}
\frac{dN}{dV dt} = \frac{dN_{\rm gal}}{dV} \dot{N}_{\rm SF} f_{Z} f_{\rm mass} f_{\rm sep},
\ee
where $d N_{\rm gal} / dV $ is the number density of galaxies; $ \dot{N}_{\rm SF}$ is the rate of stars formed per galaxy per unit time; $f_Z$ is the fraction of stars formed at metallicities of interest; $f_{\rm mass}$ is the fraction of stars formed in binaries in the mass range of interest; and $f_{\rm sep}$ is the fraction of binaries in the required range of separations.  We proceed to estimate the terms:

\begin{itemize}

\item{$d N_{\rm gal} / dV$} The space density of Milky Way equivalent galaxies (MWEGs) is $\sim 0.01$ Mpc$^{-3}$ \citep[e.g.,][]{ratesdoc}.

\item{$\dot{N}_{\rm SF}$} Either using the Milky Way's current star formation rate of a few $M_\odot$ yr$^{-1}$ as a proxy, or dividing an MWEG mass by a Hubble time and decrementing the result by a further factor of a few to account for the drop in the star formation rate in the nearby Universe relative to the peak at $z\sim 2$ (see \autoref{SFR}), set $\dot{N}_{\rm SF} \sim 2$ yr$^{-1}$.

\item{$f_Z$} The fraction of star formation in the Universe at metallicity $Z \leq 0.004$ at $z=0$ is only about $3\%$ (see \autoref{cosmo}), but when integrated over cosmic time (see \autoref{SFtotal}), $f_Z \sim 0.1$ of star formation occurred at $Z \leq 0.004$.

\item{$f_{\rm mass}$} When drawing from the Kroupa initial mass function (IMF), $\sim 0.03\%$ of primary stars have a mass above $60\, M_\odot$, where a significant window exists for chemically homogeneous evolution in binaries (see \autoref{parspace}).  Assuming that all massive stars have companions, if the secondary is drawn from a distribution that is flat in the mass ratio, roughly a third of such primaries will also have companions in the $\gtrsim 40\, M_\odot$ range.  Thus, we assume $f_{\rm mass} \sim 10^{-4}$.

\item{$f_{\rm sep}$} \autoref{parspace} suggests that the parameter space for homogeneous evolution, between binaries that are so compact that stars would overflow the Roche lobe at zero age on the main sequence and those which are too wide to have the rapid rotation required for Case M evolution, is roughly a factor of 2 in period.  The total range in period spans perhaps 5 orders of magnitude, with orbital separations ranging from a few solar radii to $O(1000)$ AU; given a moderate observed preference for shorter periods over a flat-in-the-log distribution, around 10\% of binaries could have initial separations in the range of interest.  Binaries that undergo Case M evolution should be sufficiently compact that most will merge after forming two black holes within a Hubble time  (e.g., supernova kicks should be small relative to the binary's orbital velocity and should not significantly impact the orbit); however, wind-driven mass loss could widen some of the binaries sufficiently to slow down the stars' rotation and bring them out of the homogeneous evolution space.  Assuming that the surviving fraction is of order unity, we set $f_{\rm sep} = 0.1$.

\end{itemize}

Substituting these terms into \autoref{drake}, we find a local merger rate estimate of
\be
\frac{dN}{dt} \sim \frac{0.01}{\textrm{Mpc}^3}  \times \frac{2}{\textrm{yr}} \times 0.1 \times 10^{-4} \times 0.1 \sim 20\ \textrm{Gpc}^{-3} \textrm{yr}^{-1}.
\ee

\section{Model Assumptions} \label{sec:inputs}

We simulate massive binary populations over cosmic time under the following assumptions.

\subsection{Initial distribution}\label{sec:initial}
We sample massive binary systems by assuming that the primary mass  $m_1$ follows a Kroupa IMF \citep{Kroupa+2003}. We adopt a flat mass ratio distribution for $q\in[0.1,1]$, where $q = m_2/m_1$ \citep[e.g.][]{Sana+2012, Kobulnicky+2014}.  For the distribution of orbital periods $P$ we adopt $dN/d\log_{10}P \propto (\log_{10} P ) ^{-0.5}$ appropriate for O-type stars \citep{Sana+2012}, where we extend the period range to $\log_{10} (P/\textrm{days})  \in [0.075, 3.5]$.   The lower limit is chosen by estimating the likely minimum of the underlying distribution consistent with the shortest-period system in the observed sample of 34 binaries with constrained orbital periods \citep{Sana+2012} having a period of 1.41 days, or  $\log_{10} (P/\textrm{days})=0.15$. The extension of the upper limit allows for wide binary stars \citep[e.g.,][]{Sana+2014} and single stars \citep[see discussion in][]{de-Mink+2015}. In practice, we are only interested in short-period binaries, and the remaining binaries are included in the simulation only for normalization. We assume that all binaries are circular, as expected for the short-period tidally locked systems of interest here.

We implicitly assume here that the distributions are separable and that the distributions measured in resolved nearby stellar populations are a fair approximation for the distribution of binary properties at higher redshift and lower metallicity. The latter assumption is consistent with the observational data available so far, which show no statistically significant trends with metallicity or environment \citep[e.g.,][]{Moe+2013, Sana+2013}. 

 \subsection {Stellar radii}\label{sec:radii}
For the stellar radii at zero-age we use a fit against zero-age main-sequence models computed with Eggleton's evolutionary code  \citep{Eggleton1971a} with updates by \citet{Pols+1995} and  \citet{Glebbeek+2008}.  To check if a star fills its Roche lobe, we compare its radius with the volume-equivalent effective radius fit of \citet{Eggleton1983}.

 \subsection {Threshold for homogeneous mixing}\label{sec:threshold}

We base our simulations on the grid of detailed models by \citet{Yoon+2006}.  These models are computed with a hydrodynamic stellar evolution code  which includes the effect of the centrifugal force on the stellar structure, chemical mixing and transport of angular momentum due to rotationally induced hydrodynamic instabilities \citep{Heger+2000}, and the transport of angular momentum due to magnetic torque \citep{Spruit2002}. 

The threshold rotation rate for a single star to undergo homogeneous evolution can be inferred from the grid of models by \citet{Yoon+2006}.  They express the threshold as a function of the ratio of the equatorial velocity to the Keplerian velocity, $\omega_c = v/v_k$, where they define $v_k = \sqrt{G m /r}$ with $m$ denoting the stellar mass and $r$ the stellar radius, ignoring possible deformation.  We have approximated the minimum $\omega_c$ for single stars to achieve rotationally-induced quasi-homogeneous evolution with the following analytic fit for the $Z=0.004$ grid.  
\begin{align*}
\omega_c = & 
\begin{cases}
   0.2+ 2.7 \times 10^{-4} \, \left(\frac{m}{M_\odot}-50\right)^2     & \text{for  } m <50 M_\odot, \\
 0.2    &\text{for  } m\ge50 M_\odot. \\
  \end{cases}
\end{align*}

\citet{Yoon+2006} produce models with step sizes of $0.1$ in $\omega_c$, and our fit lies close to the maximum $\omega_c$ for single stars to evolve on the standard evolutionary tracks.  On the other hand, mixing in stars in a tidally locked binary is likely to be stronger than in single stars with the same initial rotational frequency due to the stronger deviations from symmetry and particularly due to the extra reservoir of orbital angular momentum which can be fed back into the stars as they evolve.   The threshold initial rotation rate for chemically homogeneous evolution in single stars is therefore probably a conservative lower limit for the threshold in tidally locked binary stars.  Future 3D hydrodynamical simulations will be needed to investigate the threshold.  We adopt the expression above as our current best guess for the threshold in binary systems. 

After drawing systems from the initial distributions, we compute the Keplerian rotation rate assuming that the stellar spin is synchronized with the orbit, which is appropriate for the short-period systems of interest \citep{Zahn1989}.  If the stellar rotation rate is larger than our threshold for chemically homogeneous evolution we follow the evolution further. 

\subsection {Mass loss and formation of the black hole remnant}\label{sec:massloss}

We account for the mass loss driven by stellar winds and by envelope ejection during the final explosion, which affects the orbital separation and the masses of the final remnants. We adopt a simple parametrized approach, which is sufficient given the approximate nature of this calculation. Since the rates are dominated by systems formed at a metallicity of $Z \leq 0.004$ we adopt typical values for this metallicity. This is a conservative assumption, since mass loss via radiatively driven stellar winds is reduced at lower metallicity \citep{Vink+2001}.

We adopt $f_{\rm MS} = 0.1$ for the fraction of mass that is lost during the main sequence evolution, and $f_{\rm WR} = 0.25$ for the fraction of mass lost during post main sequence evolution as a Wolf-Rayet star. These are consistent with the results of \citet{Yoon+2006}, taking into account that their mass loss rates are over-predicted by their assumption that the dredge-up of CNO is as effective as Fe.  

Both the mass fallback fraction \citep[e.g.,][]{Fryer1999a, Belczynski:2012,Fryer:2012} and black-hole natal kicks \citep[e.g,][]{Repetto:2012,Janka:2013,MillerJones:2014} accompanying BH-forming core collapse supernovae are highly uncertain.  We adopt $f_{\rm SN} = 0.1$ for the fraction of mass lost during the supernova explosion, consistent with the low mass loss predicted by \citet{Fryer+2011} and \citet{Fryer:2012}.  The natal kicks are not expected to be important given the very compact binaries considered here: even the high natal kicks of $\sim 400$ km/s claimed by \citet{RepettoNelemans:2015} for the most extreme BHs  are lower than the typical orbital velocity of $\sim 800$ km/s. Moreover, evidence for such high kicks is disputed by \citet{Mandel:2015kicks} and \citet{Belczynski:2015}, who show that the existing observations are consistent with much lower natal kicks of $\lesssim 100$ km/s.   Therefore, we ignore BH natal kicks in our analysis.

We account for the possibility that the most massive helium stars end their lives as pair-instability supernovae and do not leave a remnant behind.  We therefore adopt an upper limit of 63$\Msun$ \citep{Heger+2002} for the final, pre-explosion mass of the star to form a black hole.  

\subsection{Orbital evolution}\label{sec:evolution}

We account for changes in the orbit due to wind-driven and supernovae mass loss in the Jeans mode approximation, i.e., assuming that the mass loss is spherical and fast compared to the orbital motion.  We assume that the specific angular momentum of mass lost in the stellar wind is equal to the orbital angular momentum of the star. This approximation is a conservative estimate which provides an upper limit to the orbital widening.  In reality the stellar winds may not be fast enough to satisfy this approximation, potentially leading to less widening and yielding more systems that can evolve homogeneously.    We account for the effect of widening due to instantaneous mass loss during the supernova, 
\begin{equation}
a_\textrm{after} = \frac{m_\textrm{after}}{2 m_\textrm{after} - m_\textrm{before}} a_\textrm{before}\, ,
\end{equation} 
but assume that the binary remains circular throughout its evolution, given that kick velocities are expected to be low relative to the orbital velocities. 

When two black holes are formed, we follow the decay of the orbit driven by energy and angular momentum loss resulting from the emission of gravitational waves.  The time to merger for a circular black hole binary with radius $a$ and component masses $m_1, m_2$ is given by \citet{Peters1964}:
\begin{equation}
\tau_\textrm{GW} = \frac{5}{256} \frac{c^5}{G^3} \frac{a^4}{m_1 m_2 (m_1+m_2)}\, .
\end{equation}
Given the short (few Myr) evolutionary timescale of very massive stars, the merger timescale dominates the total time between star formation and merger, so we set the time delay between formation and merger equal to $\tau_\textrm{GW}$.

\subsection{Cosmology}\label{cosmo}

\begin{figure}\center
  \includegraphics[width=\columnwidth]{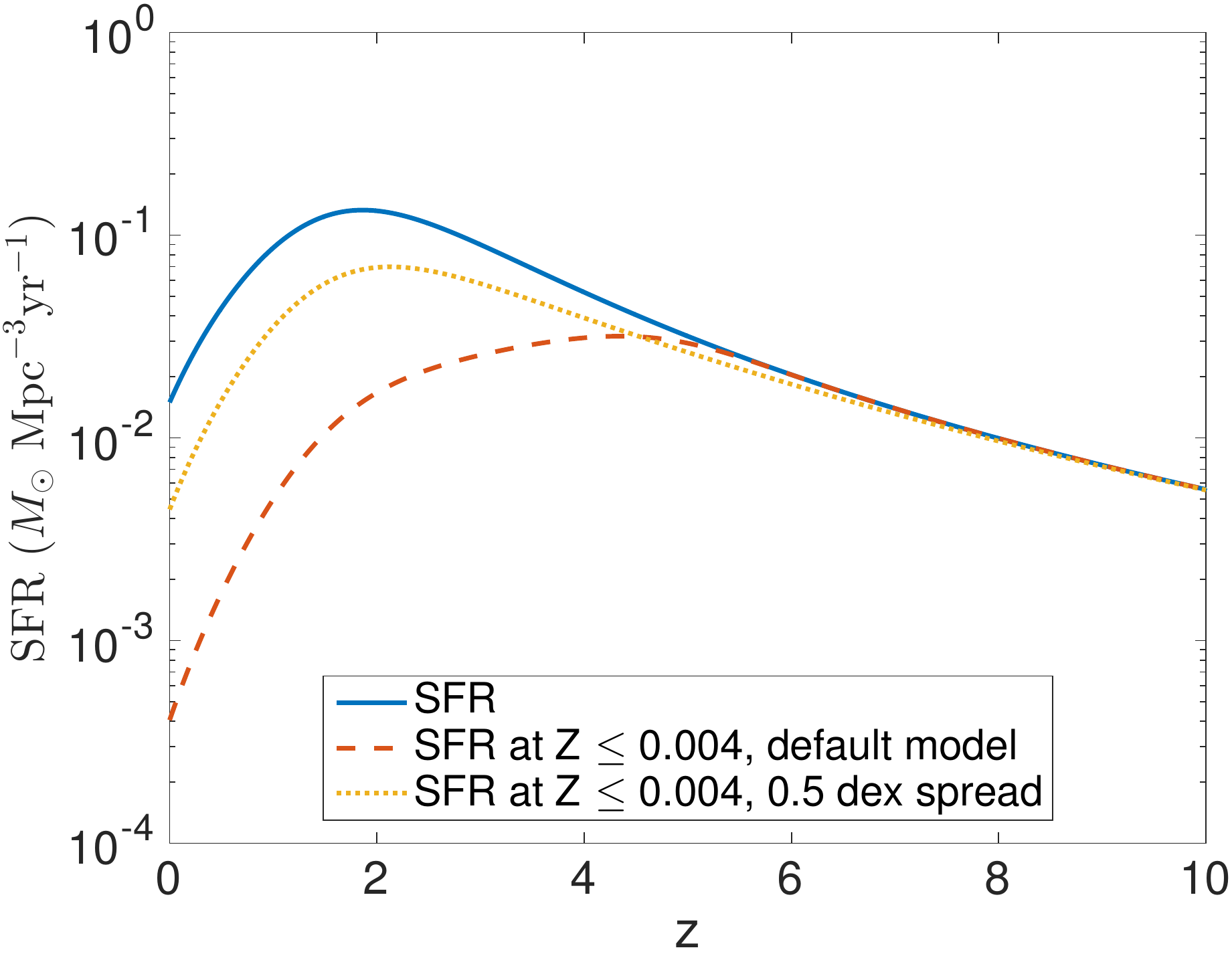}
  \caption{Total star formation rate (SFR) \citep[solid blue;][]{MadauDickinson:2014}, star formation rate at metallicity $Z \leq 0.004$ \citep[dashed red;][]{LangerNorman:2006} and star formation rate at $Z \leq 0.004$ for an alternative model with a $0.5$ dex spread in metallicity at each redshift (dotted yellow) as a function of redshift.}
   \label{SFR}
\end{figure}

We use the cosmological model of WMAP-9 \citep{WMAP9}.  We follow \citet{MadauDickinson:2014} [see their Eq.~(15)] in modeling the star formation rate $M_\textrm{SFR}$ per unit source time per unit comoving volume as a function of redshift $z$ as
\begin{equation}\label{SFReq}
\frac{d^2 M_\textrm{SFR}}{ dt dV_c} = 0.015 \frac{(1+z)^{2.7}}{[1+(1+z)/2.9]^{5.6}} \, \frac{M_\odot}{\textrm{Mpc}^{3} \textrm{yr}}\, .
\end{equation}

For the metallicity distribution as a function of redshift, we use the fit of \citet{LangerNorman:2006} based in turn on the mass--metallicity relation of \citet{Savaglio:2005} and the average cosmic metallicity scaling of \cite{KewleyKobulnicky:2005,KewleyKobulnicky:2007}.  The fraction of star formation occurring at metallicity $\leq Z$ at redshift $z$ is
\begin{equation}\label{CDFeq}
\textrm{CDF}(Z,z) = \hat{\Gamma}\left(\alpha+2, (Z/Z_\odot)^\beta 10^{0.15 \beta z}\right),
\end{equation}
where $\alpha = -1.16$, $\beta = 2$ and $\hat{\Gamma}$ is the incomplete gamma function.  We take the solar metallicity to be $Z_\odot=0.0134$ \citep{Asplund:2009}.  \autoref{CDFeq} corresponds to a mean metallicity of $\langle Z(z) \rangle \sim 1.06 \times 10^{-0.15 z} Z_\odot $ with a standard deviation of $\sim 0.38 \langle Z \rangle$, although it should be recognized that these models have significant uncertainty.  We assume that the IMF does not depend on redshift or metallicity.  

\begin{figure}\center
  \includegraphics[width=\columnwidth]{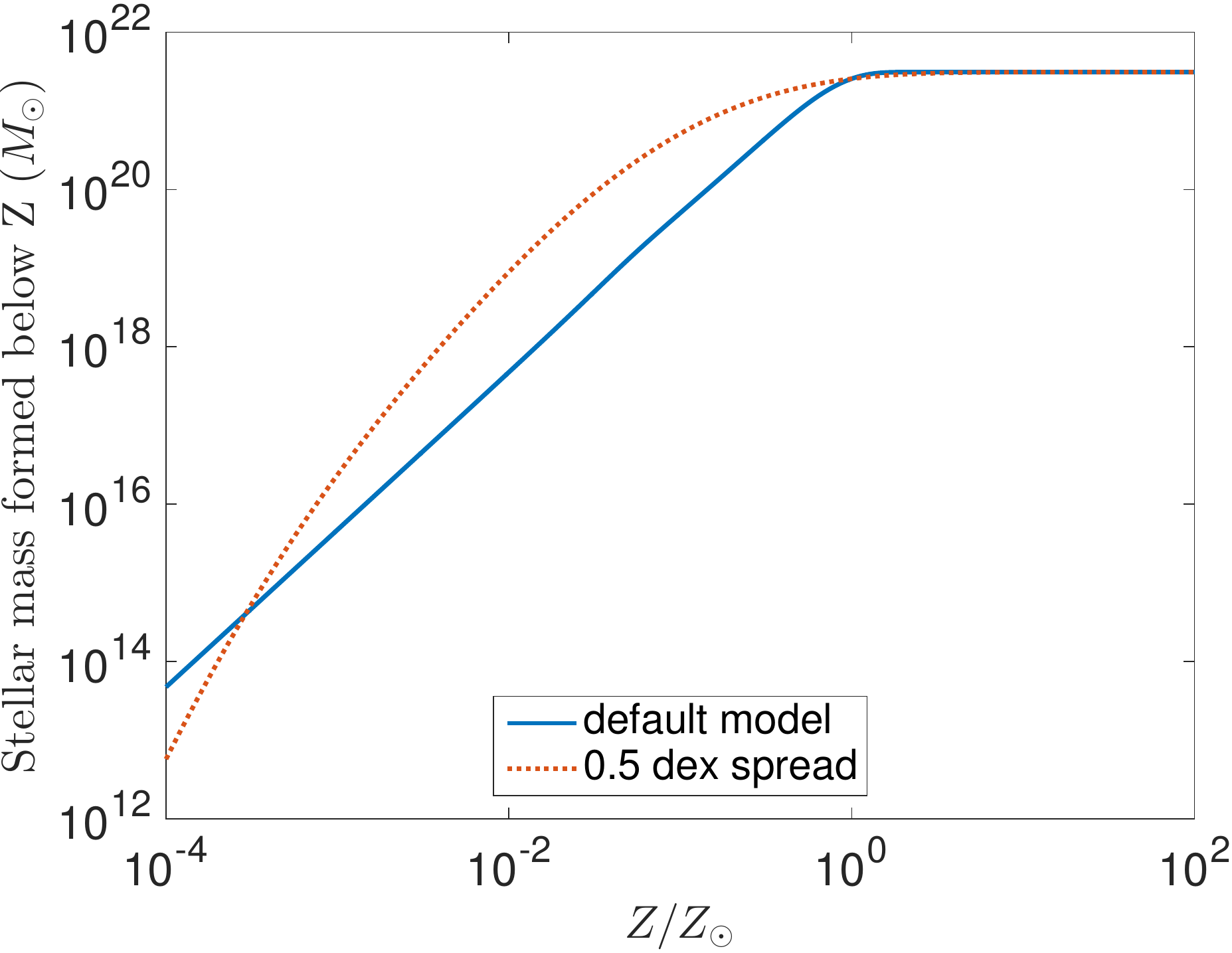}
  \caption{Cumulative stellar mass formed in the Universe at metallicities below $Z$ via the prescriptions of \citet{MadauDickinson:2014} and \citet{LangerNorman:2006} (solid blue) and via the alternative model with a $0.5$ dex spread in metallicity (dotted red).}
   \label{SFtotal}
\end{figure}

In Figure \ref{SFR} we show the total star formation rate and the star formation rate at metallicity below $Z=0.004$ as given by Eqs.~(\ref{SFReq}),(\ref{CDFeq}), while \autoref{SFtotal} shows the metallicity distribution of the total star formation in the Universe integrated over all redshifts.  The figures also show an alternative metallicity distribution $\textrm{CDF}(Z,z)$ which has the same mean metallicity at a given redshift but a broader spread of $0.5$ dex around the mean at each redshift; we use this alternative, which yields a greater low-metallicity local star formation rate, to analyze the impact of uncertainty in the metallicity distribution in \autoref{sec:uncertain}.

\section{Monte Carlo Simulation} \label{sec:MC}

We wish to estimate the rate of binary black hole mergers via the chemically homogeneous channel in tidally locked binaries, and the properties of the merging systems.

The rate of black-hole binary mergers with component masses $m_1$ and $m_2$ at the moment of merger $t_{\rm m}$ per unit source time and per unit comoving volume $V_c$ is given by:
\begin{widetext}
\be \label{eq:mergerrate}
\frac{d^4 N_{\rm merge}}{dV_{\rm c} \, dt \, dm_1 \, dm_2} (t_{\rm m})= \int_{P_{\min}}^{P_{\max}} dP  \int_0^1 dZ  \int_0^{t_{\rm m}} dt \, p(t_{\rm m}; m_1, m_2, P, Z, t_{\rm b}) \,\, \frac{d^2M_{\rm SFR}}{dt \, dV_{\rm c}} (t_{\rm b}) \,\, \frac{d^5 N_\textrm{binaries}}{dm_1 \, dm_2 \, dP \, dZ \, dM_{\rm SFR}} (t_{\rm b}) \, .
\ee
\end{widetext}

Here, $ {d^2M_{\rm SFR}}/(dt \, dV_{\rm c})$ is the star formation rate per unit time per unit comoving volume $V_{\rm c}$, evaluated at the binary birth time $t_{\rm b}$ and  ${d^5 N_\textrm{binaries}}/({dm_1 \, dm_2 \, dP \, dZ \, dM_{\rm SFR}})$ is the number density of binaries formed per unit $m_1$, $m_2$, initial orbital period $P$, and metallicity $Z$ per unit star formation rate.  The masses $m_1$ and $m_2$ refer to the black hole masses, and will differ from the birth stellar masses.  The probability density of a binary formed with given $m_1,m_2,P,Z$ at time $t_{\rm b}$ merging at time $t_{\rm m}$ is given by $p(t_{\rm m}; m_1, m_2, P, Z, t_{\rm b})$.  The innermost integral is taken over all birth times  $t_{\rm b}$ preceding the merger time $t_{\rm m}$, where the zero of time corresponds to  the Big Bang.  

The final merger rate $d^2 N_{\rm merge} / dV_\textrm{c} / dt$ is obtained by integrating \autoref{eq:mergerrate} over both component masses.  In practice, we evaluate all of these integrals with a Monte Carlo simulation.  The Monte Carlo simulation consists of two main steps, which we describe in detail below.

\subsection{Binary simulation}

We can make the calculation of \autoref{eq:mergerrate} efficient by taking advantage of the assumption that the distribution of the initial properties of a binary (IMF, period distribution) do not depend on birth time (redshift) or metallicity.  Therefore, we can simulate a set of binaries and then distribute them across cosmic time. 

At a given metallicity (we use $Z=0.004$ throughout in lieu of an integral over metallicity), we determine $d^4 N_\textrm{binaries} / (d\tau \,dm_1 \,dm_1 \, dM_{\rm SFR})$, the number of binaries per unit component mass, per unit time delay $\tau \equiv t_\textrm{m} - t_\textrm{b}$, per unit star-forming mass, by drawing $\gtrsim 10^7$ binaries with different component masses and initial periods from the initial distribution functions given in \autoref{sec:initial}.  The number of simulated binaries is chosen such that the uncertainty on the merger rate estimates from statistical fluctuations is no more than a few percent, as determined by bootstrapping.  We evolve all binaries through mass loss on the main sequence, during the helium burning stage, and during supernova, as discussed in \autoref{sec:massloss}, meanwhile evolving the binary's orbital separation to account for the mass loss (see \autoref{sec:evolution}). For future analysis, we keep only binaries that satisfy all of the following conditions:
\begin{itemize}
\item Both components have initial masses between $20$ and $300$ solar masses, which allows for all systems of interest with a very safe margin (see Figures \ref{parspace} \& \ref{survival}).  
\item Neither companion is overflowing its Roche Lobe at zero age on the main sequence (minimum initial orbital separation; see \autoref{sec:radii}).  
\item The binary is sufficiently compact to satisfy the conditions for homogeneous mixing both at the beginning and at the end of the main sequence (see \autoref{sec:threshold}).
\item Both companions have a pre-supernova mass below $63\, M_\odot$, to avoid pair-instability supernovae (see \autoref{sec:massloss}).
\item The time delay between binary formation and merger through gravitational-wave radiation reaction is less than a Hubble time (see \autoref{sec:evolution}).
\end{itemize}

The binaries of interest satisfy all of the conditions above. They provide us with a set of samples from the distribution $d^4 N_\textrm{binaries} / (d\tau \, dm_1 \, dq_1  \,  dM_{\rm SFR})$.  We normalize by the total mass of all generated binaries, $dM_{\rm SFR}$.  Each sample binary (we will label them with an index $k$ in the next section) has a formation rate of 1 per $dM_{\rm SFR}$ of star formation.  

\subsection{Merger rate calculation}

We divide the history of the Universe into a large number of bins by redshift.  Equivalently, these correspond to bins of lookback time, which we express in terms of redshift \citep[e.g.,][]{Hogg:1999} using a standard flat cosmology with $\Omega_\Lambda = 0.718$ and $h_0=0.697$ \citep{WMAP9}.  

The birth rate for a given sample binary $k$ as defined above in a given redshift bin $z_i$ is given by 
\be 
\frac{dN_{k,i}^{\rm birth}}{dt \, dV_{\rm c}}=\textrm{CDF}(Z,z_i) \, \, \frac{d^2 M_{\rm SFR}}{dt \, dV_{\rm c}} \,\, \frac{1}{dM_{\rm SFR}},
\ee
where $\textrm{CDF}(Z,z_i)$ is the fraction of star formation occurring at the metallicity of interest, i.e., at $Z \leq 0.004$, at redshift $z_i$ (see \autoref{cosmo}).

The merger rate for this sample binary $k$ in redshift bin $z_j$ is then
\be
\frac{dN_{k,j}^{\rm merge}}{dt \, dV_{\rm c}} = \sum \frac{dN_{k,i}^{\rm birth}}{dt \, dV_{\rm c}} \, \delta_{t_i+\tau_k, t_j} \, \frac{dt_i}{dt_j},
\ee
where we sum only over those birth time bins $t_i$ for which the birth time summed with the time delay between formation and merger falls into the merger time bin $t_j$ ($\delta$ is the usual Kronecker delta symbol), and the last term is included to account for differences between the time durations of different redshift bins.

\section{Results}\label{sec:results}

\begin{figure}\center
  \includegraphics[width=\columnwidth]{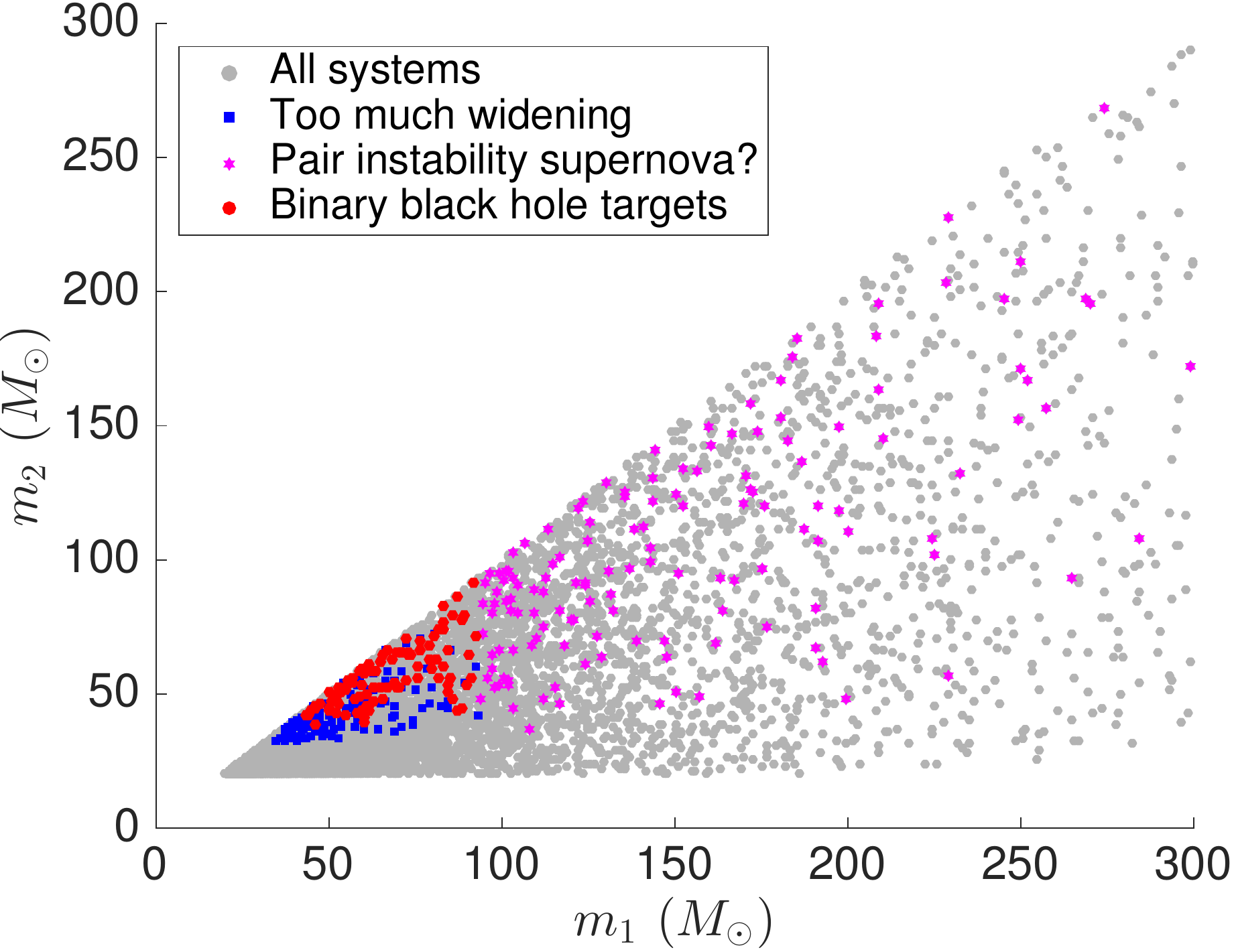}
  \caption{A representative subset of the simulated systems.  Initial masses of the simulated massive binary systems showing all chosen simulated systems (grey dots), systems that initially evolve homogeneously but widen too much during the main sequence evolution as a result of mass loss (blue squares), systems in which at least one component explodes in a pair instability supernova (magenta star) and finally systems that result in BH-BH mergers (red circles).  
   \label{survival}}
\end{figure}

Our default model predicts that there are $10.5\pm0.5$ local ($z=0$) binary black hole mergers per Gpc$^3$ per year originating from the chemically homogeneous evolution scenario. The error bar corresponds exclusively to the numerical uncertainty of the Monte Carlo integral, and does not include the systematic uncertainties in the assumed model, which are discussed in the next section. 

\autoref{survival} shows the population of binaries with initial component masses between $20$ and $300\, M_\odot$ in our Monte Carlo simulation.  The majority of these binaries are on orbits that are initially too wide to enable chemically homogeneous evolution. A smaller subset are so close that they are already Roche-lobe overflowing at the start of the main sequence, and likely to rapidly merge.  Only about 1900 binaries out of a total of $10^8$ simulated binaries, comprising a total star-forming mass of $6\times10^7\, M_\odot$, satisfy the initial conditions for Case M evolution at zero age on the main sequence. 

Of these binaries, a subset of 700 systems widen so much following mass loss on the main sequence that they no longer satisfy the conditions for homogeneous mixing in our default model. This mostly affects systems on the lower-mass end of the spectrum, where the initial period window for Case M evolution is quite narrow, so moderate amounts of mass loss and associated binary widening can close the window.  We probably over estimate the widening in our default model as discussed in \autoref{winds}, and some of these systems may in fact contribute to the formation of binary black hole mergers through the Case M scenario (see \autoref{widening}). 

Another 700 binaries, particularly those at higher masses, disqualify because at least one companion exceeds our threshold for exploding as a pair-instability supernova, leaving no remnant.  

\begin{figure}\center
  \includegraphics[width=\columnwidth]{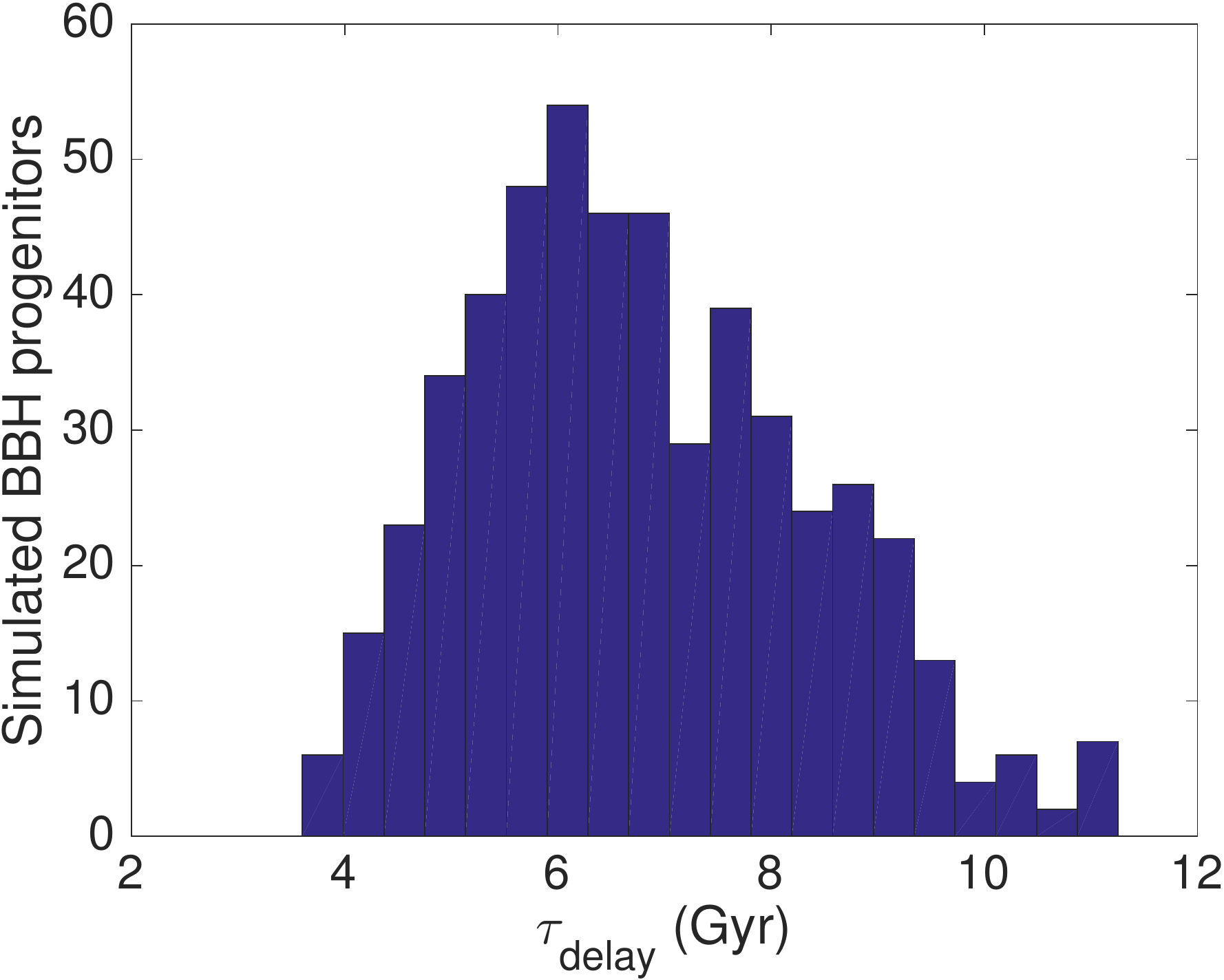}
  \caption{The distribution of delay times between formation and merger for binary black holes formed in the Case M scenario. \label{timedelay}}
\end{figure}

We find that around 500 binaries, out of $10^8$ simulated, satisfy homogeneous mixing conditions and form two merging black holes through the Case M scenario.  All mergers happen within a Hubble time.  The typical time delay between formation and merger ranges from about $4$ to $11\Gyr$, as can be seen in  \autoref{timedelay}.  

There are $8.5$ delayed binary black hole mergers per $10^6\, M_\odot$ of star formation.   For comparison, this would correspond to a``Milky-Way equivalent rate'' of $30$ mergers per Myr following the definition of  \citet{Dominik+2012} and \citet{de-Mink+2015} if all star formation occurred at $Z \leq 0.004$.  This rate refers to a Galactic steady-state rate using the normalization of $3.5\, M_\odot$ per year of Galactic star formation.

\subsection {Cosmic merger rate}

The present-day local ($z=0$) merger rate is approximately $10^{-8}$ Mpc$^{-3}$ yr$^{-1}$ or $10$ Gpc$^{-3}$ yr$^{-1}$.   At the highest redshifts the Universe is still too young to produce binary black holes mergers. The minimum time delay we find in our default simulation is $\sim 3.5$ Gyr.  As a result we do not find mergers at redshifts beyond  $z \sim 1.6$. The merger rate rises over cosmic time as mergers with longer delay times start to contribute, before dropping again at the present age of the Universe as the low-metallicity star formation rate decreases, leading to a peak of $\sim 20$ Gpc$^{-3}$ yr$^{-1}$ at $z\lesssim 0.5$.  

\begin{figure}\center
  \includegraphics[width=\columnwidth]{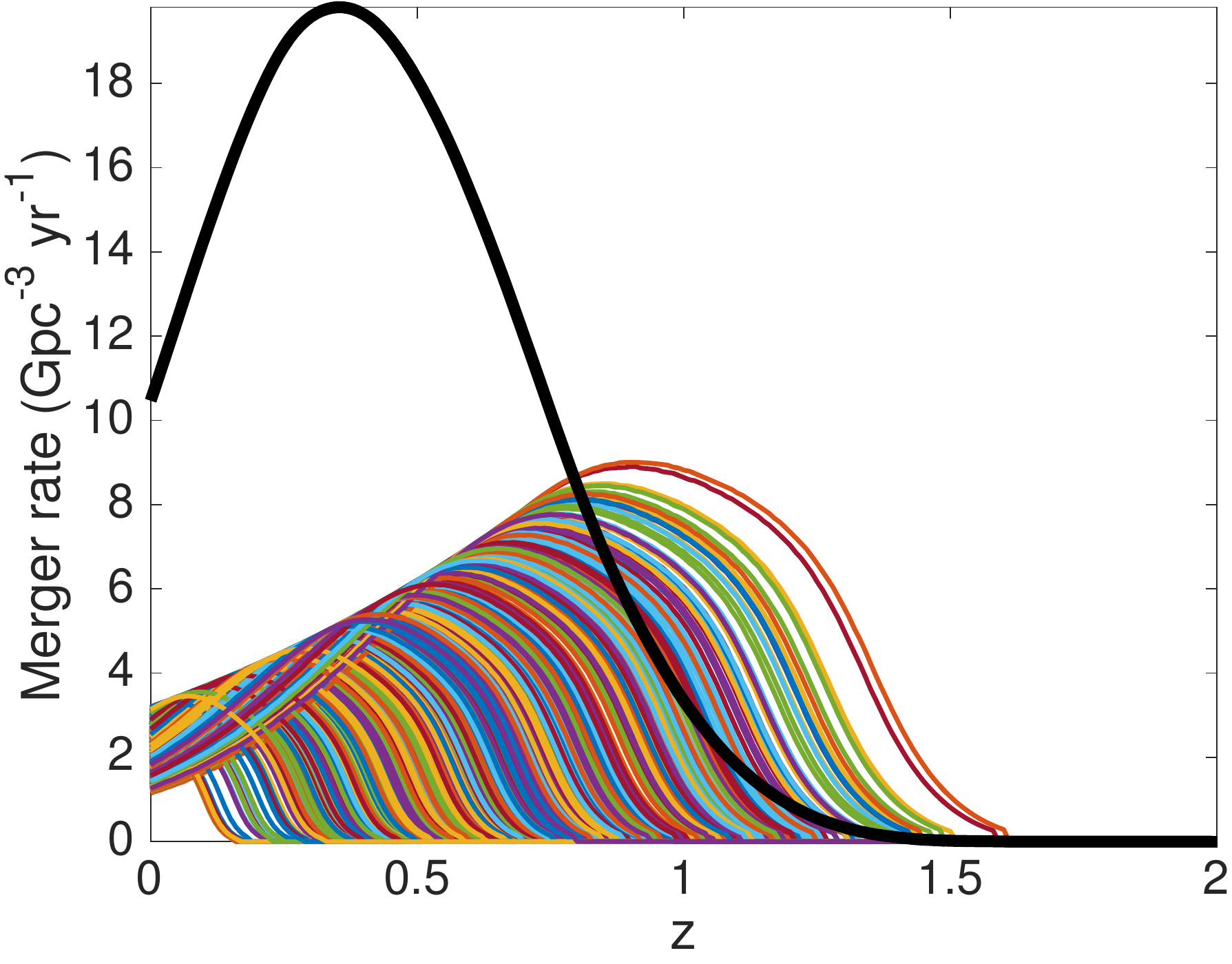}
  \caption{The total merger rate as a function of redshift is shown as a solid black line.  Colored lines show the individual contributions (multiplied by a factor 100 to fit in this plot) of the roughly 500 Case M binary black holes that have been used to populate the simulated Universe at each redshift. These curves are convolutions of the low-metallicity star formation rate with the delay time for each simulated binary.
    \label{mergerrate}}
\end{figure}

In \autoref{mergerrate} we show the total rate of mergers per year of source time per Mpc$^3$ of comoving volume as a function of redshift as a solid black line.  This line indicates the total Case M binary black hole merger rate in our simulated Universe.  We also show the individual contribution of the roughly 500 sample binary systems used to populate the simulated Universe with colored lines (rates increased by a factor of 100 for plotting); these curves are convolutions of the low-metallicity star formation rate with the delay time for each binary system.  

\begin{figure}\center
  \includegraphics[width=\columnwidth]{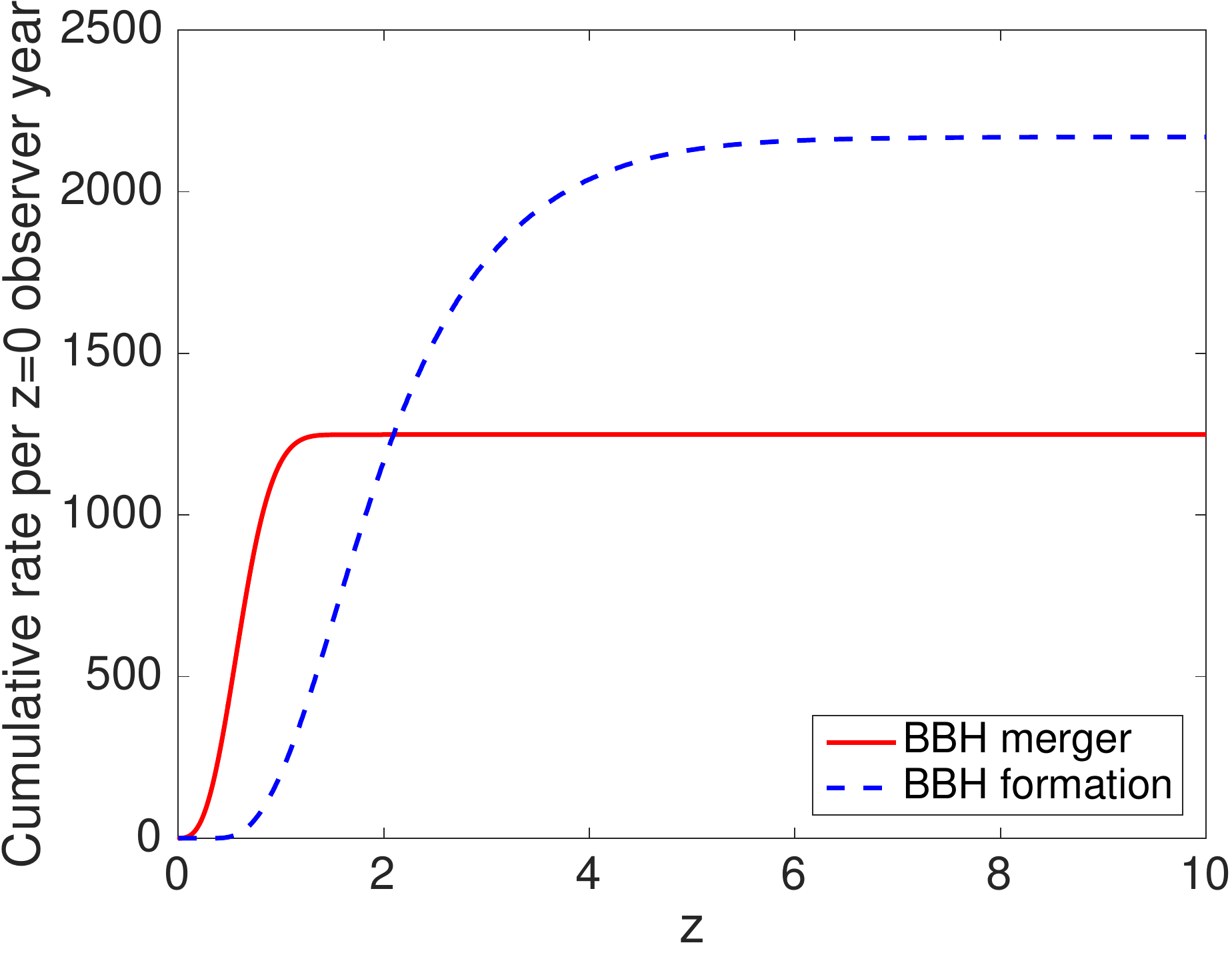}
  \caption{Cumulative merger rate (solid red) and formation rate (dashed blue) of Case M binary black holes in the Universe up to the specified redshift $z$, per year as measured by an observer at $z=0$. \label{mergersZ}}
\end{figure}

We illustrate the redshift distribution of both merger and formation rates in \autoref{mergersZ}.  We define the relevant event rate per year in the Universe up to redshift $z_\textrm{max}$ as measured by an observer at $z=0$ as
\begin{equation}
\frac{dN}{dt}(z_\textrm{max}) = \int_0^{z_\textrm{max}} \frac{d^2N}{dt\, dV_\textrm{c}} \frac{dV_\textrm{c}}{dz} \frac{1}{1+z} dz.
\end{equation} 
The rate $dN/dt\, (z_\textrm{max} \to \infty)$ then corresponds to the number of formation or merger events in the entire Universe that a present-day perfectly sensitive Earth-based detector would observe in one year.  A total of $\sim 1250$ binary black holes formed through the chemically homogeneous evolution channel merge in the Universe per year of local ($z=0$) observer time.  The median formation redshift, $1.9$, is much larger than the median merger redshift, $0.6$, because of the significant time delays between formation and merger.  The cumulative formation rate is higher than the cumulative merger rate as some of the formed binaries will only merge in the future from the perspective of the local observer.

\subsection{Merger properties}

\begin{figure}\center
  \includegraphics[width=\columnwidth]{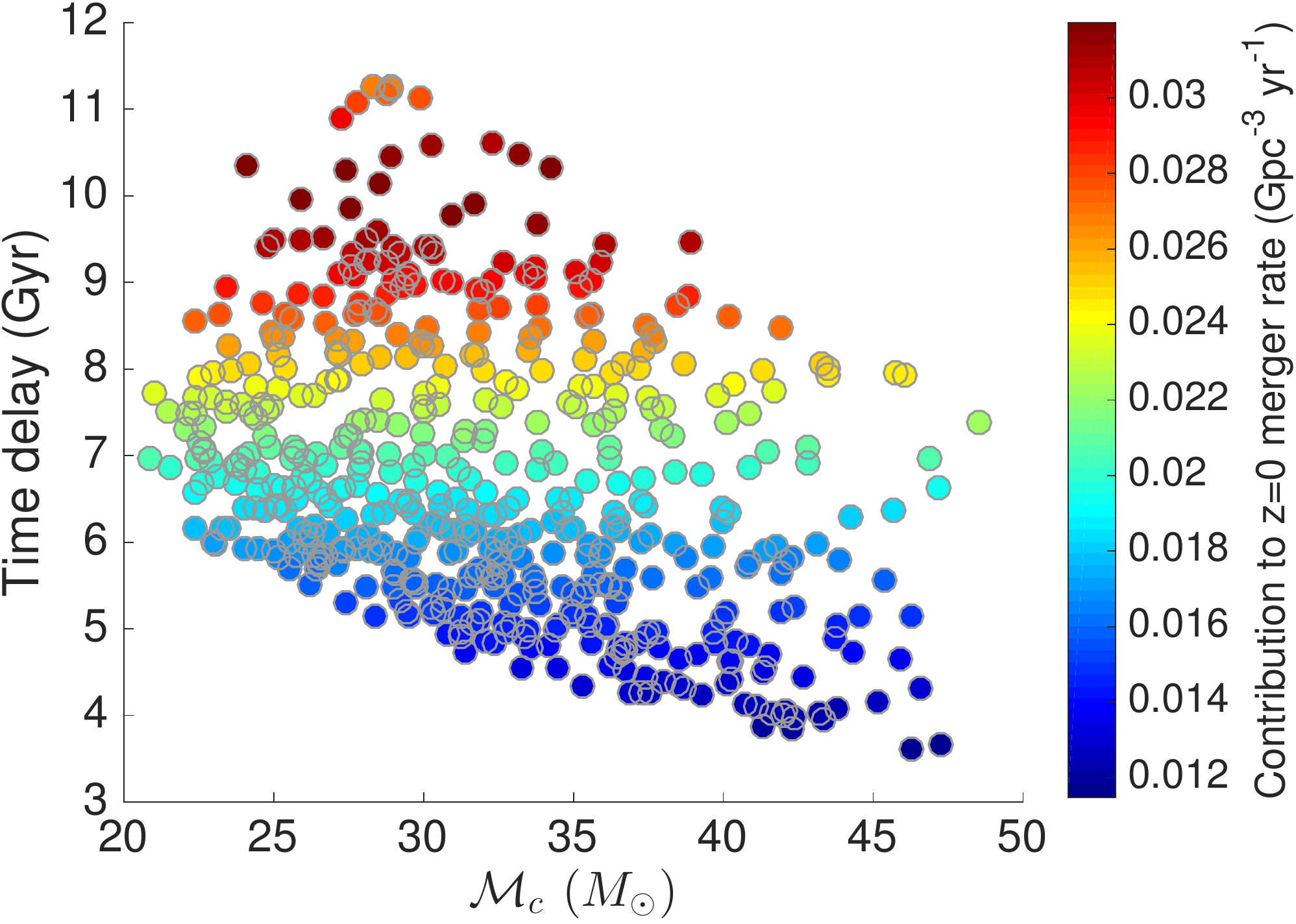}
  \caption{Contribution of individual Monte Carlo simulated binaries to the local merger rate, by chirp mass and time delay.    \label{mergersMc}}
\end{figure}

\autoref{mergersMc} displays the distribution of time delays and chirp masses $\mathcal{M}_c \equiv m_1^{3/5} m_2^{3/5} (m_1+m_2)^{-1/5}$ of the binary back holes sampled in the Monte Carlo simulation.  The size and color of each symbol indicate the local rate of mergers contributed by the given simulated binary.  Because much of the binary black hole formation occurs at redshifts $z \gtrsim 2$, binaries with longer time delays contribute more to the local merger rate.  On the other hand, more massive binaries tend to have shorter time delays, since they take less time to evolve from a fixed orbital separation through gravitational-wave emission.

\begin{figure}\center
\vspace{-1.25in}
\hspace{-0.2in}
  \includegraphics[width=0.52\textwidth]{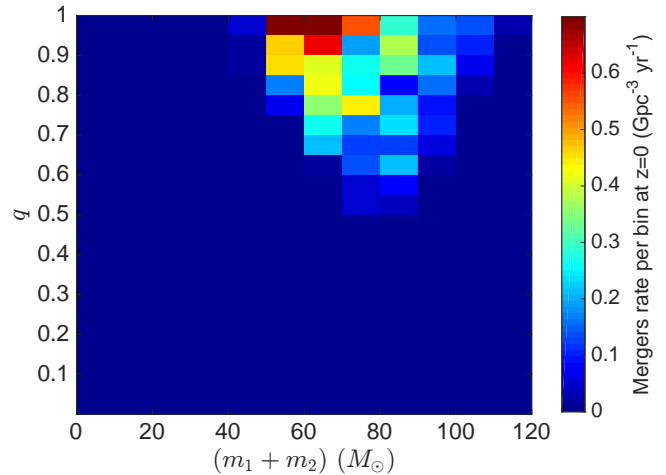}
\vspace{-1.2in}  
  \caption{Distribution of black hole binaries merging at $z=0$ in total mass $(m_1+m_2)$ and mass ratio $q=m_2/m_1$.  The color denotes the rate of mergers in each bin of width $10\, M_\odot$ in total mass space and height $0.05$ in $q$ space. \label{Mtotq} }
\end{figure}

In \autoref{Mtotq} we show the distribution of the mass ratio $q=m_2/m_1$ and the total mass  $(m_1+m_2)$ of binary black holes merging at $z=0$.  Our simulations predict typical values of $m_1+ m_2 \sim 50$ -- $110 \Msun$.  The mass distribution shows greater support for high masses than either classical population-synthesis predictions for field binary black holes \citep[e.g.,][]{Dominik+2015} or dynamically formed binary black hole models in globular clusters \citep[e.g.,][]{Rodriguez:2015}.  There is a strong preference for comparable mass ratios; there are no binaries of interest with $q<0.5$ and $70$\% of mergers come from sources with $q>0.75$.

\begin{table*}
\caption{Cosmic merger rate in our default model and the impact of various uncertainties. \label{tabres}}
\begin{center}
\begin{tabular}{lccccl}
\hline \hline
Simulation  & $R_{\rm local}$ &  $R_{\max}$ & $z(R_{\max})$ & Description & Comment \\
		  & (Gpc$^{-3}$yr$^{-1}$) &  (Gpc$^{-3}$yr$^{-1}$) &  & & \\
\hline
Default    &   10    & 20  & 0.5         & Sect. 4         & Standard simulation\\
Alternative 1    &   3   & 10  & 0.5  &  Sect. 7.1  & Reduced Case M window  \\
Alternative 2.1    &   2   & 3.5  & 0.3  & Sect. 7.1 &  Reduced Metallicity threshold ($Z \leq 0.002$) \\
Alternative 2.2   &  15 & 30 & 0.3 & Sect. 7.1 & Relaxed metallicity threshold ($Z \leq 0.008$) \\ 
Alternative 3.1  &     7    &  200 &  2 &  Sect. 7.2 & Slow winds (fixed separation)  \\
Alternative 3.2   &     7    &  500 &  4 &  Sect. 7.2 & Slow winds (halving separation)   \\
Alternative 4.1  &     -    &  - &  - &  Sect. 7.3 & Enhanced mass loss (doubled) \\
Alternative 4.2  &     25    &  50 &  0.4 &  Sect. 7.3 & Enhanced mass loss \& slow winds \\
Alternative 4.3 &     5    &  500 &  3 &  Sect. 7.3 & Reduced mass loss (by factor of 5)   \\
Alternative 5     &   10    & 20  & 0.5 & Sect. 7.4 & Increased PISN threshold ($80\Msun$) \\
Alternative 6   &   80    & 10  & 0.15 & Sect. 7.5 & Enhanced metallicity spread (0.5 dex) \\
\hline
\end{tabular}
\end{center}
\end{table*}

\section{Discussion of uncertainties in the model} \label{sec:uncertain}

Our predictions for binary black hole mergers originating from the chemically homogeneous formation scenario are affected by several major uncertainties.  The predictions are not sensitive to the poorly understood Roche-lobe overflow and common-envelope phases, which are crucial in the standard isolated binary black hole formation scenario.  They are, however, quite sensitive to the uncertain internal mixing processes and several other model assumptions.  We highlight the main ones below.  A summary is given in Table~\ref{tabres}.

\subsection{Conditions for chemically homogeneous evolution}\label{altCHE}

Whether mixing processes in rotating stars are efficient enough to trigger chemically homogeneous evolution is a matter of debate (see \autoref{CHEsingle}).  There are hints coming from observations, but they remain inconclusive so far (see \autoref{observations}). The model predictions all arise from 1D models, whose predictive power in the regime of these rapid rotation rates is limited.  Different models of rotating stars have been produced by following different theoretical frameworks and assumption. They differ, for example, in whether the mixing processes are treated as a diffusive process or whether the advective nature of meridional circulation is accounted for  \citep[cf.][]{Endal+1976, Zahn1992}. They also differ in whether or not angular momentum transport by magnetic fields is included.  As a general trend, the conditions for homogeneous evolution are obtained more easily, i.e., at lower rotation rates, when the Spruit--Tayler dynamo \citep{Spruit2002} is adopted, which is the case in the models by \citet{Yoon+2006} on which we have based our simulations.

After the submission of this work, a study by \citet{Marchant+2016} became available, exploring the parameter space for chemically homogeneous evolution with the MESA code \citep{Paxton+2015}. The window for chemically homogeneous evolution obtained by \citet{Marchant+2016} is shifted to shorter periods (they include systems that start as over-contact binaries at zero-age) and higher masses.  \citet{Marchant+2016} also find a stronger preference for lower metallicity.

The threshold for chemically homogeneous evolution depends on several uncertain assumptions: in particular the role of the mean molecular weight gradient in inhibiting the mixing processes and the inclusion of other processes for internal angular momentum transport such as by internal magnetic fields.  The window for this scenario in the initial binary parameter space and its metallicity dependence is therefore very uncertain. It should be questioned whether the window opens up at all.  

As one model alternative, we consider a more conservative fit through the data of \citet{Yoon+2006} than described in \autoref{sec:threshold} (Model Alternative 1 in Table~\ref{tabres}).  In this variation, we set the minimum $\omega_c$ for chemically homogeneous evolution to 
\begin{align*}
\omega_c = & 
\begin{cases}
   0.25+ 3.2 \times 10^{-4} \, \left(\frac{m}{M_\odot}-46\right)^2     & \text{for  } m <46 M_\odot, \\
 0.25    &\text{for  } m\ge46 M_\odot. \\
  \end{cases}
\end{align*}
This more stringent requirement on the minimal rotational frequency roughly halves the parameter space window for the Case M scenario (see the dotted line in \autoref{parspace}).   As a result, the local $z=0$ merger rate drops to $\sim 3$ Gpc$^{-3}$ yr$^{-1}$, with a peak merger rate of only $\sim 10$ Gpc$^{-3}$ yr$^{-1}$ at $z \sim 0.5$.  The binaries which satisfy these more stringent requirements typically merge in $\sim 4$ -- $7$ Gyr, as wider binaries with longer time delays which satisfied the requirements of \autoref{sec:threshold} for chemically homogeneous evolution no longer do so.  Merging binary black holes have nearly equal masses, with the mass ratio $q \gtrsim 0.7$ for all surviving systems.

On the other hand, we may have been conservative by basing our estimates on models of single stars. Stars in close binary systems may experience additional mixing processes that are not accounted for in the simulations.  We also ignored a possible contribution of systems that temporarily fill their Roche lobes or even evolve into a shallow contact phase. Such systems will likely shrink; tides will then force both stars to spin faster, which will likely enhance the internal mixing processes. If the mixing processes trigger chemically homogeneous evolution the stars may shrink within their Roche lobes. They may detach and recover from the shallow contact phase. Such systems can possibly still contribute to the formation of double black hole binaries. 

We also consider varying the maximal metallicity threshold for chemically homogeneous evolution from the default-model value of $Z=0.004$, while keeping the mass loss rate prescription unchanged (Model Alternative 2.1).  If the threshold is lowered to $Z \leq 0.002$, the binary black hole merger rate decreases to $\sim 2$ Gpc$^{-3}$ yr$^{-1}$ locally, and $\sim 3.5$ Gpc$^{-3}$ yr$^{-1}$ at its peak at $z \sim 0.3$.  Meanwhile, relaxing the chemically homogeneous evolution threshold to $Z \leq 0.008$ (Model Alternative 2.2) allows a greater fraction of binaries to evolve through the Case M channel, increasing the local merger rate to $\sim 15$ Gpc$^{-3}$ yr$^{-1}$, with a peak of $\sim 30$ Gpc$^{-3}$ yr$^{-1}$ at $z \sim 0.3$.

\subsection{Wind-driven orbital evolution}\label{widening}

For the default model, we adopted the simple assumption of fast stellar winds, i.e., Jeans-mode mass loss.  By doing so, we are likely to over predict the amount of widening as a result of mass loss (see \autoref{winds}). For the close binaries considered here, the companion lies within the stellar wind acceleration zone and the velocities of the components relative to each other are comparable to the terminal wind velocities.  This means that gravitational interaction of the winds and the stars can in principle drain additional energy and angular momentum from the system, leading to tighter binary systems (see \autoref{winds}).  We therefore consider two simple variations on our default assumptions.   

In the first alternative, we keep the binary's orbital separation fixed throughout the evolution, independently of the mass loss rates (Model Alternative 3.1).  Because the systems stay more compact, most binaries that initially satisfy the criterion for homogeneous mixing continue to do so throughout their evolution: roughly twice as many binaries per unit star-forming mass satisfy the condition for chemically homogeneous evolution at the end of the main sequence as in the default model.  On the other hand, the formed black hole binaries have shorter time delays to merger ($\sim 0.5$ -- $2$ Gyr) as the binaries do not expand in response to mass loss.  Consequently, the merger rate is much higher at larger redshifts, with a maximum of $\sim 200$ Gpc$^{-3}$ yr$^{-1}$ at $z \sim 2$, but drops to only $\sim 7$ Gpc$^{-3}$ yr$^{-1}$ locally.

In the second alternative, we halve the binary's semimajor axis by the end of the evolution to account for extreme angular-momentum dissipation by slow winds (Model Alternative 3.2).  We assume that homogeneously mixed stars shrink sufficiently during their evolution that Roche lobe overflow at later evolutionary stages is not a concern despite the binary's hardening.  In this alternative model, the same simulated systems survive as in the previous one; however, the time delays are significantly shorter for these extremely tight binaries, with all binary black holes merging within $< 0.15$ Gyr.   Even though this further increases the overall rate of mergers over cosmic time, with $\sim 2 \times 10^5$ mergers per local observer year throughout the Universe, most of these happen at higher redshifts.  The merger rate peaks at a maximum of $\sim 500$ Gpc$^{-3}$ yr$^{-1}$ at $z \sim 4$, but again drops to only $\sim 7$ Gpc$^{-3}$ yr$^{-1}$ locally.

\subsection{Mass loss through stellar winds and during the supernovae}

Given the large uncertainty in mass loss rates (see \autoref{winds}), we consider two extreme variations. First, we double mass loss rate prescriptions at all stages of the evolution: on the main sequence, during helium burning, and during the core-collapse supernovae (Model Alternative 4.1). With the default prescription for angular momentum loss (which over estimates the widening), this widens the binaries so much that few are still homogeneously mixed at the end of the main sequence. Under this alternative assumption, none merge within a Hubble time.   

If we repeat the enhanced mass loss simulation above but keep the orbital separation constant to account for angular momentum loss by slow winds (Model Alternative 4.2), we find that twice as many binaries survive as in the default model. This is in part because more initially massive binaries avoid the fate of pair instability supernovae. At the same time the fraction of binaries that are too wide to be homogeneously evolving at the end of the main sequence drops because the binary's orbit does not expand.  On the other hand, the high mass loss means that the merging binaries have a lower mass: the minimal total binary mass is only $\sim 25\, M_\odot$, while the mass ratio reaches down to $q \gtrsim 0.3$ for this alternative model, with time delays spanning $\sim 3$ -- $14$ Gyr. The net effect is to increase the local merger rate to $\sim 25$ Gpc$^{-3}$ yr$^{-1}$, with a peak of nearly $\sim 50$ Gpc$^{-3}$ yr$^{-1}$ at $z \sim 0.4$.

We also consider reducing the mass loss by a factor of 5, within the fast wind assumption (Model Alternative 4.3). With such small mass loss, the binary does not significantly expand during evolution, appreciably reducing the fraction of binaries that do not satisfy the homogeneous evolution condition at the end of the main sequence and reducing the time delays for binaries of interest to $<1$ Gyr. At the same time, a higher fraction of potentially interesting binaries run afoul of the pair instability constraint, and the surviving binaries are slightly more massive than in the default model. The merger rate peaks at $\sim 500$ Gpc$^{-3}$ yr$^{-1}$ at $z\sim 3$, but the local merger rate is only $\sim 5$ Gpc$^{-3}$ yr$^{-1}$, half that of the default model, because of the short time delays between star formation and binary merger.

\subsection{Pair instability supernovae}

To investigate the impact of the uncertain mass threshold for pair instability supernovae, we consider a variation of the criterion for the onset of a pair instability supernova (Model Alternative 5). We change the mass cutoff from the default value $\geq63\, M_\odot$ (see \autoref{sec:massloss}) to $\geq 80\, M_\odot$.  This results in a slightly larger total number of binary black holes, particularly adding more massive systems which would have exploded as pair instability supernovae under the default model, up to a total mass of $\sim 130\, M_\odot$ for merging binary black holes. The merger rate is similar to the default model within Monte Carlo numerical uncertainty.

\subsection{Metallicity distribution\label{metallicity}}

The metallicity distribution of star-forming gas in the Universe is poorly observationally constrained \citep{MadauDickinson:2014}.  As a possible alternative to the default model described in \autoref{cosmo}, we considered a metallicity distribution with the same mean metallicity as a function of redshift as in the default model, but a broader spread of $0.5$ dex around the mean at each redshift (Model Alternative 6).  This broadening of the metallicity distribution increases the fraction of local star formation that occurs in low-metallicity environments (see \autoref{SFR}).  Consequently, the peak merger rate is  $\sim 100$ Gpc$^{-3}$ yr$^{-1}$ at $z \sim 0.15$, with the local merger rate only slightly lower, $\sim 80$ Gpc$^{-3}$ yr$^{-1}$.

\subsection{Other uncertainties}

Other uncertainties, which we have not specifically modeled here, include initial conditions, such as the binary period and mass ratio distributions and the initial mass function.  These have been considered in the context of standard isolated binary evolution modeling by \citet{de-Mink+2015}, who found that the initial conditions affected the overall rate normalization but had relatively little impact on the merging binary properties.
 
\section{Conclusions} \label{sec:summary}

We have considered the evolution of close massive binary stars that could give rise to binary black hole mergers through a new evolutionary channel.  In sufficiently massive tight binaries the mixing processes triggered by rotation and tides can potentially cause the stars to evolve chemically homogeneously, leading to contraction during the evolution and preventing Roche lobe overflow \citep[][]{de-Mink+2009a}.  

We show that such systems can give rise to a significant rate of binary black hole mergers at redshifts $0 \leq z \lesssim 1.5$, peaking at $\sim 20$ Gpc$^{-3}$ yr$^{-1}$ at $z \sim 0.5$ in our default simulations.   This new channel is competitive, in terms of rates, with the classical channels such as those forming binary black holes from wider binaries which require shrinking during a common-envelope phase \citep[e.g.,][]{Dominik+2015} and dynamically-formed binary black holes in globular clusters \citep[e.g.,][]{Rodriguez:2015}.  

The predicted merger rate is consistent with the lack of detections during initial LIGO-Virgo runs, which placed upper limits of $\sim 70$ -- $170$ Gpc$^{-3}$ yr$^{-1}$ in the mass bins of interest \citep{HighMass:S6}, above our predicted local ($z=0$) merger rate of $\sim 10$ Gpc$^{-3}$ yr$^{-1}$.  It is also within the range of binary black hole merger rate predictions given in \citet{ratesdoc}, a factor of two above the ``realistic'' rate quoted there (note, however, that the scenario analyzed here yields relatively massive black holes, whose coalescence will be accompanied by loud gravitational-wave signals).

We predict that the merging binary population arising from this channel is characterized by nearly equal black hole masses and high total binary masses (typical masses of $\sim 50$--110 $M_\odot$).  Possible supernova natal kicks are expected to be small in comparison to the high orbital velocities of the progenitor stars. We therefore expect the black-hole spins to be nearly aligned, if the black hole spin directions are conserved during supernovae (i.e., there are no spin tilts).   These features could be used to observationally distinguish a population of such homogeneously evolved binaries \citep[see, e.g.,][for a discussion of clustering on gravitational-wave observations to search for subpopulations]{Mandel:2015}.  

In the standard isolated binary evolution channel, binaries are significantly hardened during the common-envelope phase leading to some mergers with very short time delays after formation, yielding a prediction of many high-redshift mergers.  In contrast, the chemically homogeneous evolution channel does not produce short time-delay mergers in our default model, with minimal delays of at least a few Gyr leading to few merging binary black holes beyond $z \sim 1.5$.   A stochastic gravitational-wave background from multiple unresolvable sources could be used to probe the existence of a population at higher redshift \citep{Mandic:2012}.  

There are many uncertainties surrounding this evolutionary channel. On the positive side, it does not suffer from the key uncertainties that plague the classical binary black hole formation channels, particularly the common-envelope phase, which is avoided in the channel discussed here.  The key uncertainties for the chemically homogeneous channel lie in the efficiency of the mixing processes in tidally locked binaries and the impact of stellar winds on orbital evolution, which can possibly close off this channel completely, but more likely change the predicted rates by factors of several.  

Electromagnetic observational constraints for this evolutionary scenario have so far proven challenging.  For example, it may in principle be possible to observe a chemically homogeneous massive star orbiting around a black hole as an intermediate stage in the evolution, which may be detectable as a high-mass X-ray binary.  However, this phase is short-lived and such observations can only be done in nearby galaxies where low-metallicity environments are rare.  This means that constraints from gravitational-wave observations, either through detections or non-trivial upper limits, will be particularly valuable for this new evolutionary scenario. 

\vspace{0.05in} 
{\bf Acknowledgements.} 

Various people independently remarked on the possible importance of the chemically homogeneous evolutionary channel in the context of binary black hole formation, including but not limited to Krzysztof Belczynski, James Guillochon and Cole Miller.  SdM acknowledges Matteo Cantiello for starting the original discussion of the possibility of this channel leading to \citet{de-Mink+2009a}, and Sung-Chul Yoon for sharing the grid of models from \citet{Yoon+2006}.  We thank Christopher Berry, Thomas Dent, Vicky Kalogera, Gijs Nelemans, Colin Norman, Abel Schootemeijer, and especially Yuri Levin for discussions and comments on the manuscript. We further thank the referee George Meynet for his suggestions. 
The authors acknowledge the Leiden Lorentz Center workshop ``The Impact of Massive Binaries Throughout the Universe''.
SdM acknowledges support by a Marie Sklodowska-Curie 
Reintegration Fellowship (H2020 MSCA-IF-2014, project id 661502).

\bibliography{my_bib,Mandel}

\begin{thebibliography}{126}
\expandafter\ifx\csname natexlab\endcsname\relax\def\natexlab#1{#1}\fi

\bibitem[{{Aasi} {et~al.}(2013){Aasi}, {Abadie}, {Abbott}, {Abbott}, {Abbott},
  {Abernathy}, {Accadia}, {Acernese}, {Adams}, {Adams}, \&
  et~al.}]{HighMass:S6}
{Aasi}, J. {et~al.} 2013, \prd, 87, 022002, 1209.6533

\bibitem[{{Aasi} {et~al.}(2015){Aasi}, {Abbott}, {Abbott}, {Abbott},
  {Abernathy}, {Ackley}, {Adams}, {Adams}, {Addesso}, \& et~al.}]{AdvLIGO}
------. 2015, Classical and Quantum Gravity, 32, 074001, 1411.4547

\bibitem[{Abadie {et~al.}(2010)}]{ratesdoc}
Abadie, J., {et~al.} 2010, Classical and Quantum Gravity, 27, 173001, 1003.2480

\bibitem[{{Abbott} {et~al.}(2016){Abbott}, {Abbott}, {Abbott}, {Abernathy},
  {Acernese}, {Ackley}, {Adams}, {Adams}, {Addesso}, {Adhikari}, \&
  et~al.}]{scenarios}
{Abbott}, B.~P. {et~al.} 2016, Living Reviews in Relativity, 19, 1304.0670

\bibitem[{Acernese {et~al.}(2015)}]{AdvVirgo}
Acernese, F., {et~al.} 2015, Class. Quant. Grav., 32, 024001, 1408.3978

\bibitem[{{Almeida} {et~al.}(2015){Almeida}, {Sana}, {de Mink}, {Tramper},
  {Soszy{\'n}ski}, {Langer}, {Barb{\'a}}, {Cantiello}, {Damineli}, {de Koter},
  {Garcia}, {Gr{\"a}fener}, {Herrero}, {Howarth}, {Ma{\'{\i}}z Apell{\'a}niz},
  {Norman}, {Ram{\'{\i}}rez-Agudelo}, \& {Vink}}]{Almeida+2015}
{Almeida}, L.~A. {et~al.} 2015, ArXiv e-prints, 1509.08940

\bibitem[{{Asplund} {et~al.}(2009){Asplund}, {Grevesse}, {Sauval}, \&
  {Scott}}]{Asplund:2009}
{Asplund}, M., {Grevesse}, N., {Sauval}, A.~J., \& {Scott}, P. 2009, \araa, 47,
  481, 0909.0948

\bibitem[{{Belczynski} {et~al.}(2013){Belczynski}, {Bulik}, {Mandel},
  {Sathyaprakash}, {Zdziarski}, \& {Miko{\l}ajewska}}]{CygnusX3:2012}
{Belczynski}, K., {Bulik}, T., {Mandel}, I., {Sathyaprakash}, B.~S.,
  {Zdziarski}, A.~A., \& {Miko{\l}ajewska}, J. 2013, \apj, 764, 96, 1209.2658

\bibitem[{{Belczynski} {et~al.}(2016){Belczynski}, {Repetto}, {Holz},
  {O'Shaughnessy}, {Bulik}, {Berti}, {Fryer}, \& {Dominik}}]{Belczynski:2015}
{Belczynski}, K., {Repetto}, S., {Holz}, D.~E., {O'Shaughnessy}, R., {Bulik},
  T., {Berti}, E., {Fryer}, C., \& {Dominik}, M. 2016, \apj, 819, 108

\bibitem[{{Belczynski} {et~al.}(2012){Belczynski}, {Wiktorowicz}, {Fryer},
  {Holz}, \& {Kalogera}}]{Belczynski:2012}
{Belczynski}, K., {Wiktorowicz}, G., {Fryer}, C.~L., {Holz}, D.~E., \&
  {Kalogera}, V. 2012, \apj, 757, 91, 1110.1635

\bibitem[{{Bethe} \& {Brown}(1998)}]{Bethe+1998}
{Bethe}, H.~A., \& {Brown}, G.~E. 1998, \apj, 506, 780, astro-ph/9802084

\bibitem[{{Bloom} {et~al.}(1999){Bloom}, {Sigurdsson}, \& {Pols}}]{Bloom+1999}
{Bloom}, J.~S., {Sigurdsson}, S., \& {Pols}, O.~R. 1999, \mnras, 305, 763,
  astro-ph/9805222

\bibitem[{{Brookshaw} \& {Tavani}(1993)}]{Brookshaw+1993}
{Brookshaw}, L., \& {Tavani}, M. 1993, \apj, 410, 719

\bibitem[{{Brott} {et~al.}(2011{\natexlab{a}}){Brott}, {de Mink}, {Cantiello},
  {Langer}, {de Koter}, {Evans}, {Hunter}, {Trundle}, \& {Vink}}]{Brott+2011}
{Brott}, I. {et~al.} 2011{\natexlab{a}}, \aap, 530, A115, 1102.0530

\bibitem[{{Brott} {et~al.}(2011{\natexlab{b}}){Brott}, {Evans}, {Hunter}, {de
  Koter}, {Langer}, {Dufton}, {Cantiello}, {Trundle}, {Lennon}, {de Mink},
  {Yoon}, \& {Anders}}]{Brott+2011a}
------. 2011{\natexlab{b}}, \aap, 530, A116, 1102.0766

\bibitem[{{Bulik} {et~al.}(2011){Bulik}, {Belczynski}, \&
  {Prestwich}}]{Bulik+2011}
{Bulik}, T., {Belczynski}, K., \& {Prestwich}, A. 2011, \apj, 730, 140,
  0803.3516

\bibitem[{{Cantiello} {et~al.}(2009){Cantiello}, {Langer}, {Brott}, {de Koter},
  {Shore}, {Vink}, {Voegler}, {Lennon}, \& {Yoon}}]{Cantiello+2009}
{Cantiello}, M. {et~al.} 2009, \aap, 499, 279, 0903.2049

\bibitem[{{Cantiello} {et~al.}(2007){Cantiello}, {Yoon}, {Langer}, \&
  {Livio}}]{Cantiello+2007}
{Cantiello}, M., {Yoon}, S.-C., {Langer}, N., \& {Livio}, M. 2007, \aap, 465,
  L29, arXiv:astro-ph/0702540

\bibitem[{{Claret}(2007)}]{Claret2007a}
{Claret}, A. 2007, \aap, 475, 1019

\bibitem[{{De Donder} \& {Vanbeveren}(2004)}]{De-Donder+2004}
{De Donder}, E., \& {Vanbeveren}, D. 2004, New Astronomy, 9, 1

\bibitem[{{de Mink} \& {Belczynski}(2015)}]{de-Mink+2015}
{de Mink}, S.~E., \& {Belczynski}, K. 2015, \apj, 814, 58, 1506.03573

\bibitem[{{de Mink} {et~al.}(2009){de Mink}, {Cantiello}, {Langer}, {Pols},
  {Brott}, \& {Yoon}}]{de-Mink+2009a}
{de Mink}, S.~E., {Cantiello}, M., {Langer}, N., {Pols}, O.~R., {Brott}, I., \&
  {Yoon}, S.-C. 2009, \aap, 497, 243, 0902.1751

\bibitem[{{de Mink} {et~al.}(2008){de Mink}, {Cantiello}, {Langer}, {Yoon},
  {Brott}, {Glebbeek}, {Verkoulen}, \& {Pols}}]{de-Mink+2008c}
{de Mink}, S.~E., {Cantiello}, M., {Langer}, N., {Yoon}, S.-C., {Brott}, I.,
  {Glebbeek}, E., {Verkoulen}, M., \& {Pols}, O.~R. 2008, in IAU Symposium,
  Vol. 252, IAU Symposium, ed. L.~{Deng} \& K.~L. {Chan}, 365--370, 0805.2544

\bibitem[{{de Mink} {et~al.}(2013){de Mink}, {Langer}, {Izzard}, {Sana}, \& {de
  Koter}}]{de-Mink+2013}
{de Mink}, S.~E., {Langer}, N., {Izzard}, R.~G., {Sana}, H., \& {de Koter}, A.
  2013, \apj, 764, 166, 1211.3742

\bibitem[{{Detmers} {et~al.}(2008){Detmers}, {Langer}, {Podsiadlowski}, \&
  {Izzard}}]{Detmers+2008}
{Detmers}, R.~G., {Langer}, N., {Podsiadlowski}, P., \& {Izzard}, R.~G. 2008,
  \aap, 484, 831, 0804.0014

\bibitem[{{Dewi} {et~al.}(2006){Dewi}, {Podsiadlowski}, \& {Sena}}]{Dewi:2006}
{Dewi}, J.~D.~M., {Podsiadlowski}, P., \& {Sena}, A. 2006, \mnras, 368, 1742,
  arXiv:astro-ph/0602510

\bibitem[{{Dominik} {et~al.}(2012){Dominik}, {Belczynski}, {Fryer}, {Holz},
  {Berti}, {Bulik}, {Mandel}, \& {O'Shaughnessy}}]{Dominik+2012}
{Dominik}, M., {Belczynski}, K., {Fryer}, C., {Holz}, D.~E., {Berti}, E.,
  {Bulik}, T., {Mandel}, I., \& {O'Shaughnessy}, R. 2012, \apj, 759, 52,
  1202.4901

\bibitem[{{Dominik} {et~al.}(2015){Dominik}, {Berti}, {O'Shaughnessy},
  {Mandel}, {Belczynski}, {Fryer}, {Holz}, {Bulik}, \&
  {Pannarale}}]{Dominik+2015}
{Dominik}, M. {et~al.} 2015, \apj, 806, 263, 1405.7016

\bibitem[{{Dufton} {et~al.}(2011){Dufton}, {Dunstall}, {Evans}, {Brott},
  {Cantiello}, {de Koter}, {de Mink}, {Fraser}, {H{\'e}nault-Brunet},
  {Howarth}, {Langer}, {Lennon}, {Markova}, {Sana}, \& {Taylor}}]{Dufton+2011}
{Dufton}, P.~L. {et~al.} 2011, \apjl, 743, L22, 1111.0157

\bibitem[{{Dufton} {et~al.}(2013){Dufton}, {Langer}, {Dunstall}, {Evans},
  {Brott}, {de Mink}, {Howarth}, {Kennedy}, {McEvoy}, {Potter},
  {Ram{\'{\i}}rez-Agudelo}, {Sana}, {Sim{\'o}n-D{\'{\i}}az}, {Taylor}, \&
  {Vink}}]{Dufton+2013}
------. 2013, \aap, 550, A109, 1212.2424

\bibitem[{{Eddington}(1925)}]{Eddington1925}
{Eddington}, A.~S. 1925, The Observatory, 48, 73

\bibitem[{{Eggleton}(1971)}]{Eggleton1971a}
{Eggleton}, P.~P. 1971, \mnras, 151, 351

\bibitem[{{Eggleton}(1983)}]{Eggleton1983}
------. 1983, \apj, 268, 368

\bibitem[{{Ekstr{\"o}m} {et~al.}(2012){Ekstr{\"o}m}, {Georgy}, {Eggenberger},
  {Meynet}, {Mowlavi}, {Wyttenbach}, {Granada}, {Decressin}, {Hirschi},
  {Frischknecht}, {Charbonnel}, \& {Maeder}}]{Ekstrom+2012}
{Ekstr{\"o}m}, S. {et~al.} 2012, \aap, 537, A146, 1110.5049

\bibitem[{{Eldridge} \& {Stanway}(2012)}]{Eldridge+2012}
{Eldridge}, J.~J., \& {Stanway}, E.~R. 2012, \mnras, 419, 479, 1109.0288

\bibitem[{{Endal} \& {Sofia}(1976)}]{Endal+1976}
{Endal}, A.~S., \& {Sofia}, S. 1976, \apj, 210, 184

\bibitem[{{Endal} \& {Sofia}(1978)}]{Endal+1978}
------. 1978, \apj, 220, 279

\bibitem[{{Evans} {et~al.}(2011){Evans}, {Taylor}, {H{\'e}nault-Brunet},
  {Sana}, {de Koter}, {Sim{\'o}n-D{\'{\i}}az}, {Carraro}, {Bagnoli}, {Bastian},
  {Bestenlehner}, {Bonanos}, {Bressert}, {Brott}, {Campbell}, {Cantiello},
  {Clark}, {Costa}, {Crowther}, {de Mink}, {Doran}, {Dufton}, {Dunstall},
  {Friedrich}, {Garcia}, {Gieles}, {Gr{\"a}fener}, {Herrero}, {Howarth},
  {Izzard}, {Langer}, {Lennon}, {Ma{\'{\i}}z Apell{\'a}niz}, {Markova},
  {Najarro}, {Puls}, {Ramirez}, {Sab{\'{\i}}n-Sanjuli{\'a}n}, {Smartt},
  {Stroud}, {van Loon}, {Vink}, \& {Walborn}}]{Evans+2011}
{Evans}, C.~J. {et~al.} 2011, \aap, 530, A108, 1103.5386

\bibitem[{{Fryer}(1999)}]{Fryer1999a}
{Fryer}, C.~L. 1999, \apj, 522, 413, arXiv:astro-ph/9902315

\bibitem[{{Fryer} {et~al.}(2012){Fryer}, {Belczynski}, {Wiktorowicz},
  {Dominik}, {Kalogera}, \& {Holz}}]{Fryer:2012}
{Fryer}, C.~L., {Belczynski}, K., {Wiktorowicz}, G., {Dominik}, M., {Kalogera},
  V., \& {Holz}, D.~E. 2012, \apj, 749, 91, 1110.1726

\bibitem[{{Fryer} \& {Heger}(2011)}]{Fryer+2011}
{Fryer}, C.~L., \& {Heger}, A. 2011, Astronomische Nachrichten, 332, 408

\bibitem[{{Georgy} {et~al.}(2011){Georgy}, {Meynet}, \& {Maeder}}]{Georgy+2011}
{Georgy}, C., {Meynet}, G., \& {Maeder}, A. 2011, \aap, 527, A52, 1011.6581

\bibitem[{{Glebbeek} {et~al.}(2008){Glebbeek}, {Pols}, \&
  {Hurley}}]{Glebbeek+2008}
{Glebbeek}, E., {Pols}, O.~R., \& {Hurley}, J.~R. 2008, \aap, 488, 1007,
  0806.0863

\bibitem[{{Grishchuk} {et~al.}(2001){Grishchuk}, {Lipunov}, {Postnov},
  {Prokhorov}, \& {Sathyaprakash}}]{Grishchuk+2001}
{Grishchuk}, L.~P., {Lipunov}, V.~M., {Postnov}, K.~A., {Prokhorov}, M.~E., \&
  {Sathyaprakash}, B.~S. 2001, Physics Uspekhi, 44, 1, astro-ph/0008481

\bibitem[{{Grudzinska} {et~al.}(2015){Grudzinska}, {Belczynski}, {Casares}, {de
  Mink}, {Ziolkowski}, {Negueruela}, {Rib{\'o}}, {Ribas}, {Paredes}, {Herrero},
  \& {Benacquista}}]{Grudzinska+2015}
{Grudzinska}, M. {et~al.} 2015, \mnras, 452, 2773, 1504.03146

\bibitem[{{Heger} {et~al.}(2000){Heger}, {Langer}, \& {Woosley}}]{Heger+2000}
{Heger}, A., {Langer}, N., \& {Woosley}, S.~E. 2000, \apj, 528, 368,
  arXiv:astro-ph/9904132

\bibitem[{{Heger} \& {Woosley}(2002)}]{Heger+2002}
{Heger}, A., \& {Woosley}, S.~E. 2002, \apj, 567, 532, arXiv:astro-ph/0107037

\bibitem[{{Hinshaw} {et~al.}(2013){Hinshaw}, {Larson}, {Komatsu}, {Spergel},
  {Bennett}, {Dunkley}, {Nolta}, {Halpern}, {Hill}, {Odegard}, {Page}, {Smith},
  {Weiland}, {Gold}, {Jarosik}, {Kogut}, {Limon}, {Meyer}, {Tucker}, {Wollack},
  \& {Wright}}]{WMAP9}
{Hinshaw}, G. {et~al.} 2013, \apjs, 208, 19, 1212.5226

\bibitem[{{Hogg}(1999)}]{Hogg:1999}
{Hogg}, D.~W. 1999, ArXiv Astrophysics e-prints, arXiv:astro-ph/9905116,
  astro-ph/9905116

\bibitem[{{Hunter} {et~al.}(2008){Hunter}, {Brott}, {Lennon}, {Langer},
  {Dufton}, {Trundle}, {Smartt}, {de Koter}, {Evans}, \&
  {Ryans}}]{Hunter+2008a}
{Hunter}, I. {et~al.} 2008, \apjl, 676, L29, 0711.2267

\bibitem[{{Hurley} {et~al.}(2000){Hurley}, {Pols}, \& {Tout}}]{Hurley+2000}
{Hurley}, J.~R., {Pols}, O.~R., \& {Tout}, C.~A. 2000, \mnras, 315, 543,
  arXiv:astro-ph/0001295

\bibitem[{{Hurley} {et~al.}(2002){Hurley}, {Tout}, \& {Pols}}]{Hurley+2002}
{Hurley}, J.~R., {Tout}, C.~A., \& {Pols}, O.~R. 2002, \mnras, 329, 897,
  arXiv:astro-ph/0201220

\bibitem[{{Izzard} {et~al.}(2004){Izzard}, {Ramirez-Ruiz}, \&
  {Tout}}]{Izzard+2004b}
{Izzard}, R.~G., {Ramirez-Ruiz}, E., \& {Tout}, C.~A. 2004, \mnras, 348, 1215,
  arXiv:astro-ph/0311463

\bibitem[{{Janka}(2013)}]{Janka:2013}
{Janka}, H.-T. 2013, \mnras, 434, 1355, 1306.0007

\bibitem[{{Kalogera} {et~al.}(2004){Kalogera}, {Kim}, {Lorimer}, {Burgay},
  {D'Amico}, {Possenti}, {Manchester}, {Lyne}, {Joshi}, {McLaughlin}, {Kramer},
  {Sarkissian}, \& {Camilo}}]{Kalogera+2004}
{Kalogera}, V. {et~al.} 2004, \apjl, 601, L179

\bibitem[{{Kewley} \& {Kobulnicky}(2005)}]{KewleyKobulnicky:2005}
{Kewley}, L., \& {Kobulnicky}, H.~A. 2005, in Astrophysics and Space Science
  Library, Vol. 329, Starbursts: From 30 Doradus to Lyman Break Galaxies, ed.
  R.~{de Grijs} \& R.~M. {Gonz{\'a}lez Delgado}, 307

\bibitem[{{Kewley} \& {Kobulnicky}(2007)}]{KewleyKobulnicky:2007}
{Kewley}, L., \& {Kobulnicky}, H.~A. 2007, in Island Universes - Structure and
  Evolution of Disk Galaxies, ed. R.~S. {De Jong}, 435

\bibitem[{{Kobulnicky} {et~al.}(2014){Kobulnicky}, {Kiminki}, {Lundquist},
  {Burke}, {Chapman}, {Keller}, {Lester}, {Rolen}, {Topel}, {Bhattacharjee},
  {Smullen}, {Vargas {\'A}lvarez}, {Runnoe}, {Dale}, \&
  {Brotherton}}]{Kobulnicky+2014}
{Kobulnicky}, H.~A. {et~al.} 2014, \apjs, 213, 34, 1406.6655

\bibitem[{{K{\"o}hler} {et~al.}(2015){K{\"o}hler}, {Langer}, {de Koter}, {de
  Mink}, {Crowther}, {Evans}, {Gr{\"a}fener}, {Sana}, {Sanyal}, {Schneider}, \&
  {Vink}}]{Kohler+2015}
{K{\"o}hler}, K. {et~al.} 2015, \aap, 573, A71, 1501.03794

\bibitem[{{Kroupa} \& {Weidner}(2003)}]{Kroupa+2003}
{Kroupa}, P., \& {Weidner}, C. 2003, \apj, 598, 1076, astro-ph/0308356

\bibitem[{{Langer} \& {Norman}(2006)}]{LangerNorman:2006}
{Langer}, N., \& {Norman}, C.~A. 2006, \apjl, 638, L63, astro-ph/0512271

\bibitem[{{Leitherer} {et~al.}(1992){Leitherer}, {Robert}, \&
  {Drissen}}]{Leitherer+1992}
{Leitherer}, C., {Robert}, C., \& {Drissen}, L. 1992, \apj, 401, 596

\bibitem[{{Lipunov} {et~al.}(1997){Lipunov}, {Postnov}, \&
  {Prokhorov}}]{Lipunov+1997}
{Lipunov}, V.~M., {Postnov}, K.~A., \& {Prokhorov}, M.~E. 1997, \mnras, 288,
  245, astro-ph/9702060

\bibitem[{{Madau} \& {Dickinson}(2014)}]{MadauDickinson:2014}
{Madau}, P., \& {Dickinson}, M. 2014, \araa, 52, 415, 1403.0007

\bibitem[{{Maeder}(1980)}]{Maeder1980}
{Maeder}, A. 1980, \aap, 90, 311

\bibitem[{{Maeder}(1987)}]{Maeder1987}
------. 1987, \aap, 178, 159

\bibitem[{{Maeder}(2000)}]{Maeder2000}
------. 2000, \nar, 44, 291

\bibitem[{{Maeder} {et~al.}(2012){Maeder}, {Georgy}, {Meynet}, \&
  {Ekstr{\"o}m}}]{Maeder+2012}
{Maeder}, A., {Georgy}, C., {Meynet}, G., \& {Ekstr{\"o}m}, S. 2012, \aap, 539,
  A110, 1201.5013

\bibitem[{{Maeder} \& {Meynet}(2000)}]{Maeder+2000a}
{Maeder}, A., \& {Meynet}, G. 2000, \araa, 38, 143, arXiv:astro-ph/0004204

\bibitem[{{Maeder} {et~al.}(2009){Maeder}, {Meynet}, {Ekstr{\"o}m}, \&
  {Georgy}}]{Maeder+2009}
{Maeder}, A., {Meynet}, G., {Ekstr{\"o}m}, S., \& {Georgy}, C. 2009,
  Communications in Asteroseismology, 158, 72, 0810.0657

\bibitem[{{Mandel}(2016)}]{Mandel:2015kicks}
{Mandel}, I. 2016, \mnras, 456, 578, 1510.03871

\bibitem[{{Mandel} {et~al.}(2015){Mandel}, {Haster}, {Dominik}, \&
  {Belczynski}}]{Mandel:2015}
{Mandel}, I., {Haster}, C.-J., {Dominik}, M., \& {Belczynski}, K. 2015, \mnras,
  450, L85, 1503.03172

\bibitem[{{Mandel} \& {O'Shaughnessy}(2010)}]{MandelOShaughnessy:2010}
{Mandel}, I., \& {O'Shaughnessy}, R. 2010, Classical and Quantum Gravity, 27,
  114007, 0912.1074

\bibitem[{{Mandic} {et~al.}(2012){Mandic}, {Thrane}, {Giampanis}, \&
  {Regimbau}}]{Mandic:2012}
{Mandic}, V., {Thrane}, E., {Giampanis}, S., \& {Regimbau}, T. 2012, Physical
  Review Letters, 109, 171102, 1209.3847

\bibitem[{{Marchant} {et~al.}(2016){Marchant}, {Langer}, {Podsiadlowski},
  {Tauris}, \& {Moriya}}]{Marchant+2016}
{Marchant}, P., {Langer}, N., {Podsiadlowski}, P., {Tauris}, T., \& {Moriya},
  T. 2016, ArXiv e-prints, 1601.03718

\bibitem[{{Martins} {et~al.}(2013){Martins}, {Depagne}, {Russeil}, \&
  {Mahy}}]{Martins+2013}
{Martins}, F., {Depagne}, E., {Russeil}, D., \& {Mahy}, L. 2013, \aap, 554,
  A23, 1304.3337

\bibitem[{{Mennekens} \& {Vanbeveren}(2014)}]{Mennekens+2014}
{Mennekens}, N., \& {Vanbeveren}, D. 2014, \aap, 564, A134, 1307.0959

\bibitem[{{Miller-Jones}(2014)}]{MillerJones:2014}
{Miller-Jones}, J.~C.~A. 2014, \pasa, 31, 16, 1401.6245

\bibitem[{{Moe} \& {Di Stefano}(2013)}]{Moe+2013}
{Moe}, M., \& {Di Stefano}, R. 2013, \apj, 778, 95, 1309.3532

\bibitem[{{Mokiem} {et~al.}(2007){Mokiem}, {de Koter}, {Vink}, {Puls}, {Evans},
  {Smartt}, {Crowther}, {Herrero}, {Langer}, {Lennon}, {Najarro}, \&
  {Villamariz}}]{Mokiem+2007a}
{Mokiem}, M.~R. {et~al.} 2007, \aap, 473, 603, 0708.2042

\bibitem[{{Narayan} {et~al.}(1991){Narayan}, {Piran}, \&
  {Shemi}}]{Narayan:1991}
{Narayan}, R., {Piran}, T., \& {Shemi}, A. 1991, \apjl, 379, L17

\bibitem[{{Nelemans}(2003)}]{Nelemans:2003}
{Nelemans}, G. 2003, in American Institute of Physics Conference Series, Vol.
  686, The Astrophysics of Gravitational Wave Sources, ed. J.~M. {Centrella},
  263--272

\bibitem[{{O'Shaughnessy} {et~al.}(2008){O'Shaughnessy}, {Kim}, {Kalogera}, \&
  {Belczynski}}]{OShaughnessy:2008}
{O'Shaughnessy}, R., {Kim}, C., {Kalogera}, V., \& {Belczynski}, K. 2008, \apj,
  672, 479

\bibitem[{{Packet}(1981)}]{Packet1981}
{Packet}, W. 1981, \aap, 102, 17

\bibitem[{{Paxton} {et~al.}(2013){Paxton}, {Cantiello}, {Arras}, {Bildsten},
  {Brown}, {Dotter}, {Mankovich}, {Montgomery}, {Stello}, {Timmes}, \&
  {Townsend}}]{Paxton+2013}
{Paxton}, B. {et~al.} 2013, \apjs, 208, 4, 1301.0319

\bibitem[{{Paxton} {et~al.}(2015){Paxton}, {Marchant}, {Schwab}, {Bauer},
  {Bildsten}, {Cantiello}, {Dessart}, {Farmer}, {Hu}, {Langer}, {Townsend},
  {Townsley}, \& {Timmes}}]{Paxton+2015}
------. 2015, \apjs, 220, 15, 1506.03146

\bibitem[{{Penny} \& {Gies}(2009)}]{Penny+2009}
{Penny}, L.~R., \& {Gies}, D.~R. 2009, \apj, 700, 844, 0905.3681

\bibitem[{{Peters}(1964)}]{Peters1964}
{Peters}, P.~C. 1964, Physical Review, 136, 1224

\bibitem[{{Pfahl} {et~al.}(2005){Pfahl}, {Podsiadlowski}, \&
  {Rappaport}}]{Pfahl+2005}
{Pfahl}, E., {Podsiadlowski}, P., \& {Rappaport}, S. 2005, \apj, 628, 343,
  astro-ph/0502122

\bibitem[{{Phinney}(1991)}]{Phinney1991}
{Phinney}, E.~S. 1991, \apjl, 380, L17

\bibitem[{{Pols} {et~al.}(1998){Pols}, {Schr\"oder}, {Hurley}, {Tout}, \&
  {Eggleton}}]{Pols+1998}
{Pols}, O.~R., {Schr\"oder}, K.-P., {Hurley}, J.~R., {Tout}, C.~A., \&
  {Eggleton}, P.~P. 1998, \mnras, 298, 525

\bibitem[{{Pols} {et~al.}(1995){Pols}, {Tout}, {Eggleton}, \&
  {Han}}]{Pols+1995}
{Pols}, O.~R., {Tout}, C.~A., {Eggleton}, P.~P., \& {Han}, Z. 1995, \mnras,
  274, 964, arXiv:astro-ph/9504025

\bibitem[{{Pols} {et~al.}(1997){Pols}, {Tout}, {Schr\"oder}, {Eggleton}, \&
  {Manners}}]{Pols+1997}
{Pols}, O.~R., {Tout}, C.~A., {Schr\"oder}, K.-P., {Eggleton}, P.~P., \&
  {Manners}, J. 1997, \mnras, 289, 869

\bibitem[{{Postnov} \& {Yungelson}(2014)}]{PostnovYungelson:2014}
{Postnov}, K.~A., \& {Yungelson}, L.~R. 2014, Living Reviews in Relativity, 17,
  3, 1403.4754

\bibitem[{{Potter} {et~al.}(2012){Potter}, {Tout}, \& {Eldridge}}]{Potter:2012}
{Potter}, A.~T., {Tout}, C.~A., \& {Eldridge}, J.~J. 2012, \mnras, 419, 748,
  1109.0993

\bibitem[{{Ram{\'{\i}}rez-Agudelo} {et~al.}(2015){Ram{\'{\i}}rez-Agudelo},
  {Sana}, {de Mink}, {H{\'e}nault-Brunet}, {de Koter}, {Langer}, {Tramper},
  {Gr{\"a}fener}, {Evans}, {Vink}, {Dufton}, \&
  {Taylor}}]{Ramirez-Agudelo+2015}
{Ram{\'{\i}}rez-Agudelo}, O.~H. {et~al.} 2015, \aap, 580, A92, 1507.02286

\bibitem[{{Ram{\'{\i}}rez-Agudelo} {et~al.}(2013){Ram{\'{\i}}rez-Agudelo},
  {Sim{\'o}n-D{\'{\i}}az}, {Sana}, {de Koter}, {Sab{\'{\i}}n-Sanjul{\'{\i}}an},
  {de Mink}, {Dufton}, {Gr{\"a}fener}, {Evans}, {Herrero}, {Langer}, {Lennon},
  {Ma{\'{\i}}z Apell{\'a}niz}, {Markova}, {Najarro}, {Puls}, {Taylor}, \&
  {Vink}}]{Ramirez-Agudelo+2013}
------. 2013, \aap, 560, A29, 1309.2929

\bibitem[{{Repetto} {et~al.}(2012){Repetto}, {Davies}, \&
  {Sigurdsson}}]{Repetto:2012}
{Repetto}, S., {Davies}, M.~B., \& {Sigurdsson}, S. 2012, \mnras, 425, 2799,
  1203.3077

\bibitem[{{Repetto} \& {Nelemans}(2015)}]{RepettoNelemans:2015}
{Repetto}, S., \& {Nelemans}, G. 2015, \mnras, 453, 3341, 1507.08105

\bibitem[{{Ribas} {et~al.}(2000){Ribas}, {Jordi}, \&
  {Gim{\'e}nez}}]{Ribas+2000}
{Ribas}, I., {Jordi}, C., \& {Gim{\'e}nez}, {\'A}. 2000, \mnras, 318, L55

\bibitem[{{Rodriguez} {et~al.}(2015){Rodriguez}, {Morscher}, {Pattabiraman},
  {Chatterjee}, {Haster}, \& {Rasio}}]{Rodriguez:2015}
{Rodriguez}, C.~L., {Morscher}, M., {Pattabiraman}, B., {Chatterjee}, S.,
  {Haster}, C.-J., \& {Rasio}, F.~A. 2015, Physical Review Letters, 115,
  051101, 1505.00792

\bibitem[{{Sana} {et~al.}(2013){Sana}, {de Koter}, {de Mink}, {Dunstall},
  {Evans}, {H{\'e}nault-Brunet}, {Ma{\'{\i}}z Apell{\'a}niz},
  {Ram{\'{\i}}rez-Agudelo}, {Taylor}, {Walborn}, {Clark}, {Crowther},
  {Herrero}, {Gieles}, {Langer}, {Lennon}, \& {Vink}}]{Sana+2013}
{Sana}, H. {et~al.} 2013, \aap, 550, A107, 1209.4638

\bibitem[{Sana {et~al.}(2012)Sana, de~Mink, de~Koter, Langer, Evans, Gieles,
  Gosset, Izzard, Le~Bouquin, \& Schneider}]{Sana+2012}
Sana, H. {et~al.} 2012, Science, 337, 444,
  http://www.sciencemag.org/content/337/6093/444.full.pdf

\bibitem[{{Sana} {et~al.}(2014){Sana}, {Le Bouquin}, {Lacour}, {Berger},
  {Duvert}, {Gauchet}, {Norris}, {Olofsson}, {Pickel}, {Zins}, {Absil}, {de
  Koter}, {Kratter}, {Schnurr}, \& {Zinnecker}}]{Sana+2014}
{Sana}, H. {et~al.} 2014, \apjs, 215, 15, 1409.6304

\bibitem[{{Savaglio} {et~al.}(2005){Savaglio}, {Glazebrook}, {Le Borgne},
  {Juneau}, {Abraham}, {Chen}, {Crampton}, {McCarthy}, {Carlberg}, {Marzke},
  {Roth}, {J{\o}rgensen}, \& {Murowinski}}]{Savaglio:2005}
{Savaglio}, S. {et~al.} 2005, \apj, 635, 260, astro-ph/0508407

\bibitem[{{Schr\"oder} {et~al.}(1997){Schr\"oder}, {Pols}, \&
  {Eggleton}}]{Schroder+1997}
{Schr\"oder}, K.-P., {Pols}, O.~R., \& {Eggleton}, P.~P. 1997, \mnras, 285, 696

\bibitem[{{Song} {et~al.}(2013){Song}, {Maeder}, {Meynet}, {Huang},
  {Ekstr{\"o}m}, \& {Granada}}]{Song+2013}
{Song}, H.~F., {Maeder}, A., {Meynet}, G., {Huang}, R.~Q., {Ekstr{\"o}m}, S.,
  \& {Granada}, A. 2013, \aap, 556, A100, 1306.6731

\bibitem[{{Song} {et~al.}(2016){Song}, {Meynet}, {Maeder}, {Ekstr{\"o}m}, \&
  {Eggenberger}}]{Song+2015}
{Song}, H.~F., {Meynet}, G., {Maeder}, A., {Ekstr{\"o}m}, S., \& {Eggenberger},
  P. 2016, \aap, 585, A120, 1508.06094

\bibitem[{{Spruit}(2002)}]{Spruit2002}
{Spruit}, H.~C. 2002, \aap, 381, 923, arXiv:astro-ph/0108207

\bibitem[{{Stancliffe} {et~al.}(2015){Stancliffe}, {Fossati}, {Passy}, \&
  {Schneider}}]{Stancliffe+2015}
{Stancliffe}, R.~J., {Fossati}, L., {Passy}, J.-C., \& {Schneider}, F.~R.~N.
  2015, \aap, 575, A117, 1501.05322

\bibitem[{{Stanway} {et~al.}(2014){Stanway}, {Eldridge}, {Greis}, {Davies},
  {Wilkins}, \& {Bremer}}]{Stanway+2014}
{Stanway}, E.~R., {Eldridge}, J.~J., {Greis}, S.~M.~L., {Davies}, L.~J.~M.,
  {Wilkins}, S.~M., \& {Bremer}, M.~N. 2014, \mnras, 444, 3466, 1408.4122

\bibitem[{{Stevenson} {et~al.}(2015){Stevenson}, {Ohme}, \&
  {Fairhurst}}]{Stevenson:2015}
{Stevenson}, S., {Ohme}, F., \& {Fairhurst}, S. 2015, \apj, 810, 58, 1504.07802

\bibitem[{{Sweet}(1950)}]{Sweet1950}
{Sweet}, P.~A. 1950, \mnras, 110, 548

\bibitem[{{Sz{\'e}csi} {et~al.}(2015){Sz{\'e}csi}, {Langer}, {Yoon}, {Sanyal},
  {de Mink}, {Evans}, \& {Dermine}}]{Szecsi+2015}
{Sz{\'e}csi}, D., {Langer}, N., {Yoon}, S.-C., {Sanyal}, D., {de Mink}, S.,
  {Evans}, C.~J., \& {Dermine}, T. 2015, \aap, 581, A15, 1506.09132

\bibitem[{{Vink} \& {de Koter}(2005)}]{Vink+2005}
{Vink}, J.~S., \& {de Koter}, A. 2005, \aap, 442, 587, arXiv:astro-ph/0507352

\bibitem[{{Vink} {et~al.}(2000){Vink}, {de Koter}, \& {Lamers}}]{Vink+2000}
{Vink}, J.~S., {de Koter}, A., \& {Lamers}, H.~J.~G.~L.~M. 2000, \aap, 362,
  295, arXiv:astro-ph/0008183

\bibitem[{{Vink} {et~al.}(2001){Vink}, {de Koter}, \& {Lamers}}]{Vink+2001}
------. 2001, \aap, 369, 574, arXiv:astro-ph/0101509

\bibitem[{{von Zeipel}(1924{\natexlab{a}})}]{von-Zeipel1924}
{von Zeipel}, H. 1924{\natexlab{a}}, \mnras, 84, 665

\bibitem[{{von Zeipel}(1924{\natexlab{b}})}]{von-Zeipel1924a}
------. 1924{\natexlab{b}}, \mnras, 84, 684

\bibitem[{{Voss} \& {Tauris}(2003)}]{Voss+2003}
{Voss}, R., \& {Tauris}, T.~M. 2003, \mnras, 342, 1169, astro-ph/0303227

\bibitem[{{Woosley}(1993)}]{Woosley1993}
{Woosley}, S.~E. 1993, \apj, 405, 273

\bibitem[{{Woosley} \& {Heger}(2006)}]{Woosley+2006}
{Woosley}, S.~E., \& {Heger}, A. 2006, \apj, 637, 914, arXiv:astro-ph/0508175

\bibitem[{{Yoon} \& {Langer}(2005)}]{Yoon+2005}
{Yoon}, S.-C., \& {Langer}, N. 2005, \aap, 443, 643, arXiv:astro-ph/0508242

\bibitem[{{Yoon} {et~al.}(2006){Yoon}, {Langer}, \& {Norman}}]{Yoon+2006}
{Yoon}, S.-C., {Langer}, N., \& {Norman}, C. 2006, \aap, 460, 199,
  arXiv:astro-ph/0606637

\bibitem[{{Zahn}(1989)}]{Zahn1989}
{Zahn}, J.-P. 1989, \aap, 220, 112

\bibitem[{{Zahn}(1992)}]{Zahn1992}
------. 1992, \aap, 265, 115

\end{thebibliography}

\label{lastpage}
\end{document}